%% file: sn-article.tex
\documentclass[pdflatex,sn-nature,iicol]{sn-jnl}

\DeclareUnicodeCharacter{2212}{-}
\usepackage{lmodern}
\usepackage{microtype}
\usepackage{graphicx}%
\usepackage{multirow}%
\usepackage{amsmath,amssymb,amsfonts}%
\usepackage[capitalise]{cleveref}
\usepackage{amsthm}%
\usepackage{mathrsfs}%
\usepackage[title]{appendix}%
\usepackage{xcolor}%
\usepackage{textcomp}%
\usepackage{manyfoot}%
\usepackage{booktabs}%
\usepackage{algorithm}%
\usepackage{algorithmicx}%
\usepackage{algpseudocode}%
\usepackage{listings}%
\usepackage{xspace}
\usepackage{csquotes}
\usepackage{fancybox}
\usepackage{hhline}
\usepackage{fontawesome}
\usepackage{trimclip}
\usepackage{fbox}
\usepackage[version=4]{mhchem}
\usepackage{cuted}
\usepackage{svg}
\usepackage{tikz}
	\usetikzlibrary{calc}
	\usetikzlibrary{positioning}
\usepackage{color}
\usepackage{soul}
\usepackage[colorinlistoftodos,prependcaption,textsize=footnotesize,backgroundcolor=orange!20]{todonotes}
\usepackage{url}
\usepackage{siunitx}
\usepackage{etoolbox} 
\patchcmd{\endabstract}{\null}{}{}{}
\DeclareSIUnit\permille{\text{\textperthousand}}

\newsavebox\foobox 
\newlength{\foodim}

\newcommand{\ad}{A_{\text{\tiny{D}}}}
\newcommand{\ap}{{A^\prime}}
\newcommand\del\partial
\newcommand\eps\varepsilon
\newcommand{\dm}{\text{\tiny{DM}}}
\newcommand{\med}{\text{\tiny{MED}}}
\newcommand{\dark}{\text{\tiny{D}}}
\newcommand{\sm}{\text{\tiny{SM}}}

\newcommand{\lohengrin}{\textsc{Lohengrin}\xspace}
\usepackage[italic]{hepnicenames}
\makeatletter\def\@shiftlen@anti@gen@bar{0mu}\makeatother
\usepackage[compat=1.1.0]{tikz-feynhand}
\definecolor{electron}{HTML}{1f77b4}
\definecolor{crimson}{HTML}{dd143c}
\definecolor{newgreen}{HTML}{009900}

\newif\iflclip
\newif\ifbclip
\newif\ifrclip
\newif\iftclip
\def\CLIP{\dimexpr\fboxrule+.2pt\relax}
\def\nulclip{0pt}
\newcommand\partbox[2]{%
  \lclipfalse\bclipfalse\rclipfalse\tclipfalse%
  \let\lkern\relax\let\rkern\relax%
  \let\lclip\nulclip\let\bclip\nulclip\let\rclip\nulclip\let\tclip\nulclip%
  \parseclip#1\relax\relax%
  \iflclip\def\lkern{\kern\CLIP}\def\lclip{\CLIP}\fi
  \ifbclip\def\bclip{\CLIP}\fi
  \ifrclip\def\rkern{\kern\CLIP}\def\rclip{\CLIP}\fi
  \iftclip\def\tclip{\CLIP}\fi
  \lkern\clipbox{\lclip{} \bclip{} \rclip{} \tclip}{\fbox{#2}}\rkern%
}
\def\parseclip#1#2\relax{%
  \ifx l#1\lcliptrue\else
  \ifx b#1\bcliptrue\else
  \ifx r#1\rcliptrue\else
  \ifx t#1\tcliptrue\else
  \fi\fi\fi\fi
  \ifx\relax#2\relax\else\parseclip#2\relax\fi
}

\newcommand{\electron}[0]{\partbox{rb}{$e^{-}$}}
\newcommand{\positron}[0]{\partbox{rb}{$e^{+}$}}
\newcommand{\proton}[0]{\partbox{rb}{$p^{\phantom{0}}$}}
\newcommand{\neutron}[0]{\partbox{rlt}{$n^{\phantom{0}}$}}
\newcommand{\aproton}[0]{\partbox{rb}{$\bar{p}^{\phantom{0}}$}}
\newcommand{\aneutron}[0]{\partbox{rlt}{$\bar{n}^{\phantom{0}}$}}
\newcommand{\pip}[0]{\partbox{rb}{$\pi^+$}}
\newcommand{\pim}[0]{\partbox{rb}{$\pi^-$}}
\newcommand{\pimp}[0]{\partbox{rb}{$\pi^\mp$}}
\newcommand{\piz}[0]{\partbox{rlb}{$\pi^0$}}
\newcommand{\etaz}[0]{\partbox{rlb}{$\eta^{\phantom{0}}$}}
\newcommand{\mup}[0]{\partbox{rl}{$\mu^+$}}
\newcommand{\mum}[0]{\partbox{rl}{$\mu^-$}}
\newcommand{\klong}[0]{\partbox{lrt}{$K_L^0$}}
\newcommand{\kshort}[0]{\partbox{l}{$K_S^0$}}
\newcommand{\kp}[0]{\partbox{}{$K^+$}}
\newcommand{\km}[0]{\partbox{}{$K^-$}}
\newcommand{\kz}[0]{\partbox{lrt}{$K^0$}}
\newcommand{\akz}[0]{\partbox{lrt}{$\bar{K^0}$}}
\newcommand{\lamsigz}[0]{\partbox{rtl}{$\varLambda/\varSigma^0$}}
\newcommand{\sigmap}[0]{\partbox{}{$\varSigma^+$}}
\newcommand{\sigmam}[0]{\partbox{}{$\varSigma^-$}}
\newcommand{\sigmapm}[0]{\partbox{}{$\varSigma^\pm$}}

\graphicspath{{figures/}}

\newcommand{\Acts}{\texttt{Acts}\xspace}
\newcommand{\Geant}{\texttt{Geant4}\xspace}
\begin{document}

\title[A Proposal for the \lohengrin Experiment to Search for Dark Sector Particles at the ELSA Accelerator]{A Proposal for the \lohengrin Experiment to Search for Dark Sector Particles at the ELSA Accelerator}


\author*[1]{\fnm{Philip} \sur{Bechtle}}\email{bechtle@physik.uni-bonn.de}

\author[1]{\fnm{Christian} \sur{Bespin}}\email{cbespin@uni-bonn.de}

\author[2]{\fnm{Dominique} \sur{Breton}}\email{breton@lal.in2p3.fr}

\author[4]{\fnm{Carlos} \sur{Orero Canet}}\email{carlos.orero@ific.uv.es}

\author[1]{\fnm{Klaus} \sur{Desch}}\email{desch@physik.uni-bonn.de}

\author[1]{\fnm{Herbi} \sur{Dreiner}}\email{dreiner@physik.uni-bonn.de}

\author[1]{\fnm{Oliver} \sur{Freyermuth}}\email{freyermuth@physik.uni-bonn.de}

\author[1,3]{\fnm{Rhorry} \sur{Gauld}}\email{rgauld@mpp.mpg.de}

\author[1]{\fnm{Markus} \sur{Gruber}}\email{gruber@physik.uni-bonn.de}

\author[4]{\fnm{C\'{e}sar} \sur{Blanch Guti\'{e}rrez}}\email{cesar.blanch@ific.uv.es}

\author[1]{\fnm{Hazem} \sur{Hajjar}}\email{s6mohajj@uni-bonn.de}

\author*[1]{\fnm{Matthias} \sur{Hamer}}\email{hamer@physik.uni-bonn.de}

\author*[1]{\fnm{Jan-Eric} \sur{Heinrichs}}\email{heinrichs@physik.uni-bonn.de}

\author[4]{\fnm{Adrian} \sur{Irles}}\email{adrian.irles@ific.uv.es}

\author[1]{\fnm{Jochen} \sur{Kaminski}}\email{kaminski@physik.uni-bonn.de}

\author[1]{\fnm{Laney} \sur{Klipphahn}}\email{s6laklip@uni-bonn.de}

\author[1]{\fnm{Hans} \sur{Kr\"uger}}\email{krueger@physik.uni-bonn.de}

\author[1]{\fnm{Michael} \sur{Lupberger}}\email{lupberger@physik.uni-bonn.de}

\author[2]{\fnm{Jihane} \sur{Maalmi}}\email{maalmi@lal.in2p3.fr}

\author[2]{\fnm{Roman} \sur{P\"oschl}}\email{poeschl@lal.in2p3.fr}

\author[1]{\fnm{Dennis} \sur{Proft}}\email{proft@physik.uni-bonn.de}

\author[1,6]{\fnm{Leonie} \sur{Richarz}} \email{leonie.c.richarz@ntnu.no}

\author[1]{\fnm{Tobias} \sur{Schiffer}}\email{schiffer@physik.uni-bonn.de}

\author[1]{\fnm{Patrick} \sur{Schw\"abig}}\email{schwaebig@physik.uni-bonn.de}

\author*[1]{\fnm{Martin} \sur{Sch\"urmann}}\email{mar-schuermann@physik.uni-bonn.de}

\author[5]{\fnm{Dirk} \sur{Zerwas}}\email{dirk.zerwas@in2p3.fr}

\affil[1]{\orgdiv{Physikalisches Institut}, \orgname{Rheinische Friedrich-Wilhelms-Universit\"at Bonn}, \orgaddress{\street{Nussallee 12}, \city{Bonn}, \postcode{53115}, \state{NRW}, \country{Germany}}}

\affil[2]{\orgdiv{IJCLab Orsay}, \orgname{CNRS/IN2P3}, \orgaddress{\street{15 rue Georges Clémenceau}, \city{Orsay}, \postcode{91405}, \country{France}}}

\affil[3]{\orgdiv{Werner-Heisenberg-Institut}, \orgname{Max-Planck-Institut für Physik (MPP)}, \orgaddress{\street{Boltzmannstraße 8}, \city{Garching}, \postcode{85748}, \state{Bavaria}, \country{Germany}}}

\affil[4]{\orgdiv{Instituto de Fisica Corpuscalar}, \orgname{Universitat de Valencia}, \orgaddress{\street{Carrer del Vatedratic Jose Beltran Martinez 2}, \city{Valencia}, \postcode{46980}, \country{Spain}}}
\affil[5]{\orgdiv{DMLab}, \orgname{Deutsches Elektronen-Synchrotron DESY, CNRS/IN2P3}, \orgaddress{\city{Hamburg}, \country{Germany}}}
\affil[6]{\orgdiv{Department of Materials Science and Engineering}, \orgname{NTNU Norwegian University of Science and Technology}, \orgaddress{\street{Sem Saelands vei 12}, \city{Trondheim}, \postcode{7034}, \country{Norway}}}


\abstract{We present a proposal for a future light dark matter search experiment at the Electron Stretcher Accelerator ELSA in Bonn: \lohengrin. It employs the fixed-target missing momentum based technique for searching for dark-sector particles. The \lohengrin experiment uses a beam of electrons that is extracted from the ELSA accelerator and that is shot onto a thin target to produce mainly Standard Model bremsstrahlung and - in rare occasions - possibly new particles coupling feebly to the electron. A well motivated candidate for such a new particle is the dark photon, a new, possibly massive gauge boson arising from a new gauge interaction in a dark sector and mixing kinetically with the Standard Model photon. The \lohengrin experiment is estimated to reach sensitivity to couplings small enough to explain the relic abundance of dark matter in various models for dark photon masses between $\sim\SI{1}{\mega\electronvolt}$ and $\sim\SI{100}{\mega\electronvolt}$.}

\maketitle
\section{Introduction}\label{sec:intro}
\input{introduction.tex}

\section{Dark Photon Production at ELSA}\label{sec:theory}
\input{darkphotonsatelsa.tex}

\section{The \lohengrin Experiment}\label{sec:lohengrin}
\input{lohengrinexperiment.tex}

\section{Expected Physics Reach}\label{sec:lohengrin:physics}
\input{ExpectedPhysicsReach.tex}

\section{Roadmap for \lohengrin}\label{sec:roadmap}
\input{roadmap.tex}

\section{Conclusion}\label{sec:conclusion}
\input{conclusion}

\bibliography{sn-article.bib}

\backmatter

\begin{appendices}

\section{Appendix}\label{secA1}

We show a table of energetically possible background events in table \ref{tab:backgrounds}. It also indicates possible background rejection strategies by indicating relevant detector components. The notation works as follows: \partbox{rtb}{$x$}: can be measured by trackers, \partbox{rlb}{$x$}: can be vetoed by measuring energy in the ECAL, \partbox{tlb}{$x$}: can be vetoed by ECAL hits, \partbox{trl}{$x$}: can be vetoed by HCAL hits. Note that the table is not exhaustive but only illustrates the most basic final state combinations.
\clearpage
\newpage
\begin{table}[ht!]
    \caption{Table of energetically possible background events considering a \SI{3.2}{\giga\electronvolt} electron beam. We also show possible rejection strategies by indicating the detector component which can be used to veto the respective particles. The notation is to be understood like the following: line on the left: can be measured by tracker, line on top: can be vetoed by measuring ECAL energy, line on the right: can be vetoed due to hits in the ECAL, line on the bottom: can be vetoed by HCAL hits. Particularly challenging channels are marked in red.\\
    $\mathcal{H}$ represents a general hadronic target, while the symbols $N$, $\mathcal{M}$ and $\mathcal{B}$ indicate nucleons, mesons and baryons, respectively. $X$ is a placeholder for multi-meson states.}
    \label{tab:backgrounds}
    \centering
    \begin{tabular}{llll}
    \toprule
    \toprule
    Process & & & \\
    \midrule 
    Elastic scattering & $e^-\mathcal{H} \rightarrow e^-\mathcal{H}$& &  \\
    Bremsstrahlung & $e^-\mathcal{H} \rightarrow e^- \mathcal{H} \gamma$ & & \\
    \midrule
    \midrule
    \multicolumn{4}{l}{Subsequent photon conversion} \\
    \midrule
        & $\gamma^\star \rightarrow \positron\,\electron$ & $\gamma^\star \rightarrow \mup\,\mum$ & \\
        & $\gamma^\star \rightarrow \piz + X $ & $\gamma^\star \rightarrow \etaz\, + X$ & \color{red}$\gamma^\star \rightarrow \kz\,\akz $\color{black}\\
        & $\gamma^\star \rightarrow \pip\,\pim $ &  & $\gamma^\star \rightarrow \kp\,\km $\\
        & $\gamma^\star \rightarrow \proton\,\aproton $ & \color{red}$\gamma^\star \rightarrow \neutron\,\aneutron $\color{black} & \\
        
    \midrule
    \midrule
    \multicolumn{4}{l}{Electro- and subsequent photonuclear processes w/ psudoscalar meson and baryon octet} \\
    \midrule
    $\gamma^\star N \rightarrow \mathcal{B}$ & $\gamma^\star p \rightarrow \proton$ & & \\
         & $\gamma^\star n \rightarrow \neutron$\color{black} & & \\
    \midrule
    $\gamma^\star N \rightarrow \mathcal{M}\,\mathcal{B}$
        & $\gamma^\star p \rightarrow\piz\,\proton $ & $\gamma^\star p \rightarrow \etaz\,\proton$ & $\gamma^\star p \rightarrow \kz\,\sigmap$\\
        & $\gamma^\star p \rightarrow \pip\,\neutron$ &  & $\gamma^\star p \rightarrow \kp\,\lamsigz$\\
        & $\gamma^\star n \rightarrow \piz\,\neutron$ & $\gamma^\star n \rightarrow \etaz\,\neutron$ & \color{red}$\gamma^\star n \rightarrow \kz\,\lamsigz$\color{black}\\
        & $\gamma^\star n \rightarrow \pim\,\proton$ &  & $\gamma^\star n \rightarrow \kp\,\sigmam$\\
    \midrule
    $\gamma^\star N \rightarrow \mathcal{M}\,\mathcal{M}\,\mathcal{B}$
        & $\gamma^\star p\rightarrow\pip\,\pim\,\proton$ & $\gamma^\star p\rightarrow\kp\,\km\,\proton$ & $\gamma^\star p\rightarrow\pip\,\kz\,\lamsigz$\color{black}\\
        & $\gamma^\star p \rightarrow \piz\,\piz\,\proton$ & $\gamma^\star p\rightarrow\kshort\,\klong\,\proton$ & $\gamma^\star p\rightarrow\piz\,\kp\,\lamsigz$\\
        & $\gamma^\star p\rightarrow \pip\,\piz\,\neutron$ && $\gamma^\star p\rightarrow\piz\,\kz\,\sigmap$ \\
        & $\gamma^\star n \rightarrow \pip\,\pim\,\neutron$ & $\gamma^\star n \rightarrow \kp\,\km\,\neutron$ & $\gamma^\star n\rightarrow\piz\,\kz\,\lamsigz$\color{black}\\
        & $\gamma^\star n \rightarrow \piz\,\piz\,\neutron$ & \color{red}$\gamma^\star n \rightarrow \kz\,\akz\,\neutron$\color{black} & $\gamma^\star n\rightarrow\pim\,\kp\,\lamsigz$\\
        & $\gamma^\star n \rightarrow \piz\,\pim\,\proton$ & & $\gamma^\star n\rightarrow\pimp\,\kz\,\sigmapm$\\
    \midrule
    $\gamma^\star N \rightarrow \mathcal{B}\,\bar{\mathcal{B}}\,\mathcal{B}$
        & $\gamma^\star p \rightarrow \neutron\,\aneutron\,\proton$ & + combinations with $\varSigma$ and $\varLambda$ baryons &\\
        & $\gamma^\star p \rightarrow \proton\,\aproton\,\proton$ &&\\
        & \color{red}$\gamma^\star n \rightarrow \neutron\,\aneutron\,\neutron$\color{black} &&\\
        & $\gamma^\star n \rightarrow \proton\,\aproton\,\neutron$ &&\\
    \bottomrule
    \end{tabular}
\end{table}




\end{appendices}


\end{document}

%% file: introduction.tex
The nature of dark matter (DM) is unexplained in the Standard Model (SM) 
of elementary particle physics. Many different models for DM have been put 
forward and some of these models have been experimentally tested. So far searches 
for DM have produced only negative or non-reproducible results (for an overview see ~\cite{Essig:2013lka,Alexander:2016aln,Chu:2011be, Gninenko:2012eq,Blumlein:2013cua,Andreas:2012mt, Graham:2021ggy}). 
Weakly interacting, massive particles (WIMPs) with a mass in the GeV-TeV range have been considered excellent candidates for DM.
However, with the direct DM search experiments  and the indirect bounds from the LHC experiments providing ever stronger limits on the WIMP parameter space, other, more complex explanations for DM are being studied more closely. One class of such models introduces either scalar or fermion DM particles that interact through a new gauge interaction, based on a spontaneously broken symmetry. The associated gauge boson can couple to the SM sector through a mechanism that is called kinetic mixing, effectively introducing a feeble interaction between the dark sector and the 
SM. 

In this paper, we propose the experiment \lohengrin\footnote{According to the \lohengrin myth, \lohengrin will disappear immediately if someone asks for his name -- or its meaning. \url{https://en.wikipedia.org/wiki/Lohengrin}} at the Electron Stretcher Accelerator, ELSA~\cite{Hillert:2006yb, ELSAHomepage} at the University of Bonn. The aim of this experiment is to employ momentum measurements of single electrons before and after a thin target to search for the disappearance of energy and momentum in so-called \enquote{dark bremsstrahlung} processes, where a \enquote{dark photon} couples to the electron and then either leaves the detector unregistered, or serves as a portal to a \enquote{dark sector} of DM particles and converts into a pair of such (invisible) DM particles. In contrast to reappearance experiments, the sensitivity of a disappearance based experiment is not limited by the conversion of the dark sector particle back into SM particles. However, the precise measurement of the visible event kinematics, the trigger, the suppression of rare electro- or photoproduction neutral hadronic backgrounds, and the detector acceptance pose challenges.

The idea for fixed-target missing momentum based searches for dark-sector particles stems from~\cite{Izaguirre:2014bca}. The fundamental predictions have been worked out 
in~\cite{Izaguirre:2015yja,Andreas:2013iba,Jaeckel:2012mjv,LDMXTheoryPaper,TheoryOverviewOfPortalInteractions}. Fundamentally, this type of experiment could also be sensitive to parts of the parameter spaces of strongly interacting massive particle 
DM~\cite{Hochberg:2014kqa,Hochberg:2015vrg,Berlin:2018tvf}, 
elastically decoupling 
DM~\cite{Kuflik:2015isi,Kuflik:2017iqs}, 
asymmetric
 DM~\cite{Petraki:2013wwa,Zurek:2013wia}, 
freeze-in
 DM~\cite{Hall:2009bx,Becker_2024,Baker_2018},
 axion-like particles (ALPs) \cite{Jaeckel:2010ni},
 $B-L$ gauge bosons \cite{Ilten:2018crw},
and sterile neutrinos as 
DM~\cite{Dodelson:1993je}.
A large variety of new physics scenarios potentially accessible with the missing momentum strategy are discussed in \cite{LDMXTheoryPaper,TheoryOverviewOfPortalInteractions}.

Numerous other experiments with a wide range of 
experimental methods share sensitivity to these models with a  
fixed-target missing momentum based search. Amongst these are beam 
dump-experiments using proton beams and derived beams like the NA62 and the NA64 experiments~\cite{NA62:2019meo,NA62:2023qyn, 
NA64DarkMatter, NA64:2019auh, NA64:2023wbi, Mongillo_2023,MiniBooNEDM:2018cxm,NA482:2015wmo}, 
also including future options like SHiP~\cite{SHiP:2020vbd}, and phenomenological 
derivations of limits on dark sector models from these experiments~\cite{Gninenko:2012eq,Blumlein:2013cua,Andreas:2012mt}. Also at electron beams, beam dump experiments are performed~\cite{HPS:2018xkw,APEX:2011dww,Merkel:2014avp}.

Overlap in the sensitivity also exists with experiments looking for direct 
detection of dark sector DM like 

FUNK~\cite{Andrianavalomahefa:2021gtu}, 
and with collider-based searches, e.g.\ at  
KLOE~\cite{KLOE-2:2018kqf},
BESIII~\cite{BESIII:2017fwv}, 
LHCb~\cite{LHCb:2017trq,LHCb:2019vmc,Gorkavenko_2024},
BaBar~\cite{BaBar:2017tiz},
Belle~II~\cite{Belle-II:2018jsg} and 
FASER~\cite{FASER:2023tle}.

In contrast to all previous beam dump experiments~\cite{NA62:2019meo,NA62:2023qyn,NA64DarkMatter,NA64:2019auh,NA64:2023wbi,MiniBooNEDM:2018cxm,NA482:2015wmo,SHiP:2020vbd,HPS:2018xkw,APEX:2011dww,Merkel:2014avp} and a large fraction of the other experiments, the strength of a disappearance based experiment like \lohengrin with its unique region of sensitivity lays in the scaling of the dark photon coupling \(\varepsilon\) to the SM. A reappearance experiment scales with \(\varepsilon^4\), while disappearance scales with \(\varepsilon^2\). 

\lohengrin at ELSA will operate at a relatively low beam energy of \SI{3.2}{\giga\electronvolt}. Other proposals for possible disappearance-based experiments like LDMX~\cite{Akesson:2018vlm,LDMX:2019gvz,Akesson:2022vza,LDMX:2023zbn} or DarkSHINE~\cite{Chen:2022liu} would operate at \SIrange{8}{16}{\giga\electronvolt}.  While ELSA can supply electrons at a rate of up to 6\,GHz with several electrons extracted from each bunch, it also offers the  capability to deliver a beam of single electrons at variable rates up to 500\,MHz with an exceptionally low relative beam energy spread of \SI{0.8}{\permille}. The former opens the possibility for true single electron measurements up into the \({\cal O}(0.5\,\mathrm{GHz})\) regime. The small beam energy spread eliminates the need for a high resolution tagging tracker, therefore reducing the amount of material in front of the target. Compared to the expected sensitivity of LDMX and DarkSHINE, due to the lower beam energy, the sensitivity of the \lohengrin experiment will cover the smaller interval for the dark photon mass of $\SI{1}{\mega\electronvolt} \lesssim m_{A'} \lesssim \SI{50}{\mega\electronvolt}$.

While the ELSA accelerator can hence deliver single electrons at a spacing of \SI{2}{\nano\second} or more, the resolution of single events at this rate poses a considerable challenge for the detectors that are used to tag the electron in the initial state and measure the trajectories and energies of particles in the final state. 

We propose the following experimental approach in order to overcome or mitigate these challenges: the target is placed in between two planes of a telescope-like tracking detector, behind which an electromagnetic and a hadronic calorimeter are stacked. The entire tracking detector is placed inside a strong magnetic field; the bending power of the magnet system is strong enough to bend the trajectories of electrons away from the calorimeter system, even electrons with minimal energy loss in the tracking detector or the target. The tracking planes before the target comprise the tag tracker, used to determine the number of electrons in the initial state. The tracking planes behind the target comprise the recoil tracker, used to determine the momentum of any charged final state particles, including the scattered electron, and to efficiently trigger the readout of the calorimeters for events with low energy electrons in the final state. The calorimeters are used as a veto system: events with a large amount of energy deposited in the electromagnetic calorimeter or a significant amount of energy deposited in the hadronic calorimeter are vetoed.

The sensitivity goal of the \lohengrin experiment is to extend to portal couplings as low as required to fully explain the relic density of DM for dark photon masses between a few MeV and tens of MeV for $10^{14} - 10^{15}$ electrons on target (EoT). With the targeted extraction rate, this can be achieved in about 100 days of beam time. 

This paper presents a feasibility study for the \lohengrin experiment. It is based on detailed theoretical calculations for the signal process and the most important backgrounds, as well as the measured performance of existing candidates for the most important detector components of the experiment. While the existing candidates fall short of some of the requirements for this high rate experiment, necessary improvements in particular for the front-end electronics of the tracking detector and the calorimeter are discussed as a basis for the further development of the \lohengrin experiment. 

The paper is organized as follows: In \cref{sec:theory} we introduce the production mechanism of dark photons at ELSA and outline the experimental strategy. \cref{sec:lohengrin} provides details on the accelerator, the beam properties, the foreseen detector components, their performance and required improvements, the trigger and data acquisition, and of the event reconstruction. \cref{sec:lohengrin:physics} discusses the projected physics reach for a baseline analysis strategy. \cref{sec:roadmap} outlines a possible roadmap for the experiment, before \cref{sec:conclusion} provides a summary.

We use the following right-handed coordinate system throughout the remainder of this paper: the direction of motion of the incoming electrons at the location of the target marks the positive z-axis. The magnetic field is assumed to point into the direction of the positive y-axis, such that the electrons are bent towards the positive x-axis in the magnetic field. The polar angle $\theta$ is defined with respect to the z-axis and can assume values between $0$ and $\pi$. The azimuthal angle $\phi$ is defined with respect to the positive x-axis and can assume values between $0$ and $2\pi$.

%% file: darkphotonsatelsa.tex
In a particular family of dark sector (DS) models, dark photons act as a portal between the SM sector and the DS.
They can be produced by shooting an electron beam with sufficient energy on a fixed target:
In these models, the dark photon is the mediator of a new U(1)$_\dark$ gauge interaction in the DS and couples to the SM sector by means of kinetic mixing with the SM photon.
The electron beam interacts with the target, emitting SM bremsstrahlung and, occasionally, a dark photon.
The rate of dark photon emission depends on the properties of the dark photon, in particular its mass and the strength of the kinetic mixing.
In this section, we will first briefly discuss the thermal history of (light) DS in general, and then turn to the special case of dark photons and how to search for them in the \lohengrin missing momentum experiment.

\subsection{Light Thermal Dark Matter}\label{sec:theory:ltdm}
A compelling and straightforward explanation of the observed DM abundance is the production by thermal freeze-out from the plasma in the early Universe~\cite{Rubakov:2017xzr}.

The traditional focus of DM searches is the possibility of the existence of weakly interacting massive particles (WIMPs) with masses near the electroweak (EW) scale.
However, repeated null results in WIMP searches suggest to extent the laboratory searches for DM also to models with lower DM particle masses. 
Such models are compatible with the thermal freeze-out mechanism, given that the DM coupling to SM particles is scaled down as well in order to recover the correct annihilation rate:
\begin{equation}
    \langle \sigma v \rangle \sim \SI{3e-26}{\centi\meter\cubed\per\second} \sim \SI{e-9}{\per\giga\electronvolt\squared}
\end{equation}
This is also referred to as the "WIMPless \mbox{miracle}"~\cite{Feng:2008ya}.

Generically, one may consider SM extensions containing not only a single particle, but a hidden sector containing both a DM candidate with mass $m_\dm$ and a mediator with mass $m_\med$ that interact through a coupling of strength $g_\dark$.
Connecting SM and DM can be realized by a portal interaction with strength $g_\sm$, which may be small for various reasons, see \cite{DarkSector2016CommunityReport} and citations therein.

Only certain regions in the parameter space spanned by $\{g_\sm, g_\dark, m_\med, m_\dm \}$ are of phenomenological interest for accelerator experiments, depending on the hidden sector masses:
For $m_\med < m_\dm$, the DM freeze out in the early Universe happens through the "secluded" annihilation, $\text{DM}+\text{DM} \rightarrow \text{MED}+\text{MED}$, such that the thermal averaged cross section scales roughly as $\langle \sigma v \rangle \sim g_\dark^4/m_\dm^2$.
This scenario does not provide a clear parameter space target for accelerator experiments, since the coupling required to produce hidden sector signals in accelerators, $g_\sm$, can be arbitrarily small to get the correct DM relic abundance.
For $m_\med > m_\dm$, the freeze out happens through "direct" annihilation with a virtual mediator, $\text{DM}+\text{DM}\: \rightarrow \text{MED}^\ast \rightarrow \text{SM}+\text{SM}$.
Then, upon taking the non-relativistic limit, $v \ll c$, and neglecting terms of order $m_e/m_\dm$, the thermal average annihilation cross section scales as
\begin{equation}
    \label{eq:th_avg_annihilation_cs}
    \langle \sigma v \rangle \approx \frac{1}{6\pi} \frac{g_\dark^2 g_\sm^2 m_\dm^2 v^2}{(m_\med^2 - 4 m_\dm^2)^2 + m_\med^2 \varGamma^2_\med} ,
\end{equation}
where $\varGamma^2_\med$ is the mediator decay width~\cite{Battaglieri:2017aum}.
If one is sufficiently far away from the resonance and if $m_\med \gg \varGamma_\med$, the cross section depends on the beyond Standard Model (BSM) parameters only through $m_\dm$ and the dimensionless combination
\begin{equation}
    \varUpsilon \equiv g_\sm^2 \frac{g_\dark^2}{4\pi} \left(\frac{m_\dm}{m_\med}\right)^4 ,
\end{equation}
since $\langle \sigma v \rangle \sim \varUpsilon / m_\dm^2$.
The DM abundance in cosmological models requires a \textit{minimum} value of the cross section in \cref{eq:th_avg_annihilation_cs}, to avoid DM overproduction.
For fixed benchmark scenarios specifying the ratio $m_\dm/m_\med$ and $g_\dark$ (or alternatively the dark fine structure constant $\alpha_\dark = g_\dark^2/4\pi$), one thus obtains lower bounds in a parameter space spanned by $\{ m_\dm , \varUpsilon \}$ that correspond to \textit{minimum} values of $g_\sm$, which is the coupling directly testable in accelerator experiments.
These bounds provide natural so called thermal relic targets for the sensitivity of light DM search experiments.
The exact form depends on the nature of the DM particle and mediator.
Note that the above reasoning applies only off resonance.
A more detailed discussion in the $m_\med > m_\dm$ regime including the resonance effects can be found in \cite{Feng:2017drg}.

\subsection{Dark Photons as Portals to a Light Dark Sector}\label{sec:theory:dp}
An important benchmark model for light DM is a hidden sector containing a so-called dark photon together with scalar or fermion DM particles.
Dark photons serve as the mediator particles mentioned in the previous section, since they hypothetically couple to both DM through a gauge coupling and to the SM through a kinetic mixing operator, as discussed in the following paragraphs.
The mediator mass from the the previous section is thus identified with the mass eigenvalue associated to the dark photon, i.e. $m_\med = m_\ap$.

We consider a simple SM extension containing a new, broken gauge symmetry $U(1)_\text{\tiny{D}}$ with associated spin-1 field $\ad^\mu$ and field strength tensor \mbox{$F^{\mu\nu}_\text{\tiny{D}} = \del^\mu \ad^\nu - \del^\nu \ad^\mu$}.
Furthermore, the $\ad^\mu$ field might couple through dark gauge interactions to a current consisting of DM scalars or fermions, $J^\mu_\text{\tiny{D}}$.
The DS extension is connected to the SM through the kinetic mixing of the hypercharge gauge boson with the new field $\ad^\mu$.
The relevant Lagrangian reads
\begin{align}
    \label{eq:lagrange}
	\mathscr{L} \supset& -\frac{1}{4} {F_\text{\tiny{D}}}_{\mu\nu}F_\text{\tiny{D}}^{\mu\nu} + \frac{1}{2} m_{\ad}^2 {\ad}_\mu \ad^\mu \nonumber \\
    &- \frac{\sin\varepsilon_Y}{2} F_\text{\tiny{D}}^{\mu\nu} B_{\mu\nu} - g_\text{\tiny{D}} {\ad}_\mu J_\text{\tiny{D}}^\mu\, ,
\end{align}
where we assume that the $U(1)_\text{\tiny{D}}$ gauge boson has acquired a mass $m_{\ad}$ through a suitable mechanism.

The kinetic mixing operator, $\propto F_\text{\tiny{D}}^{\mu\nu} B_{\mu\nu}$, serves as portal interaction and will induce new couplings of the SM fermions to the dark photon mass eigenstate $A'_\mu$ with mass $m_{A^\prime} \approx m_{\ad}$.
At energies well below the EW scale, the SM charged fermions $f$ receive couplings to the dark photon field of the form\footnote{In fact, the kinetic mixing $\propto F_\text{\tiny{D}}^{\mu\nu} B_{\mu\nu}$ also induces nonzero dark photon couplings to neutrinos after EWSB, since the dark photon mass eigenstate will be a linear combination containing one of the $SU(2)_L$ gauge bosons \cite{Dreiner:2013mua}.}
\begin{equation}
    \label{eq:lepton_dp_interaction}
    \mathscr{L} \supset \sum_f i Q_f\varepsilon \, \bar{f} \gamma^\mu f A^\prime_\mu ,
\end{equation}
where the sum runs over all SM fermions with electric charge $Q_f$ and $\varepsilon = \varepsilon_Y \, \cos\theta_W$ is the reduced kinetic mixing parameter.
In light of the previous section we can identify the effective portal couplings $g_\sm^f = Q_f \varepsilon$ arising from this model.
Within this simple low-energy theory, the kinetic mixing parameter is arbitrary, but depending on the more fundamental theory, it can range between several orders of magnitude, $\varepsilon_Y \sim \varepsilon \sim 10^{-1} - 10^{-12}$ \cite{Fabbrichesi:2020wbt}.
For our purposes it is thus sufficient to treat it as free parameter.

\begin{figure}
    \centering
    \includegraphics[width=\columnwidth]{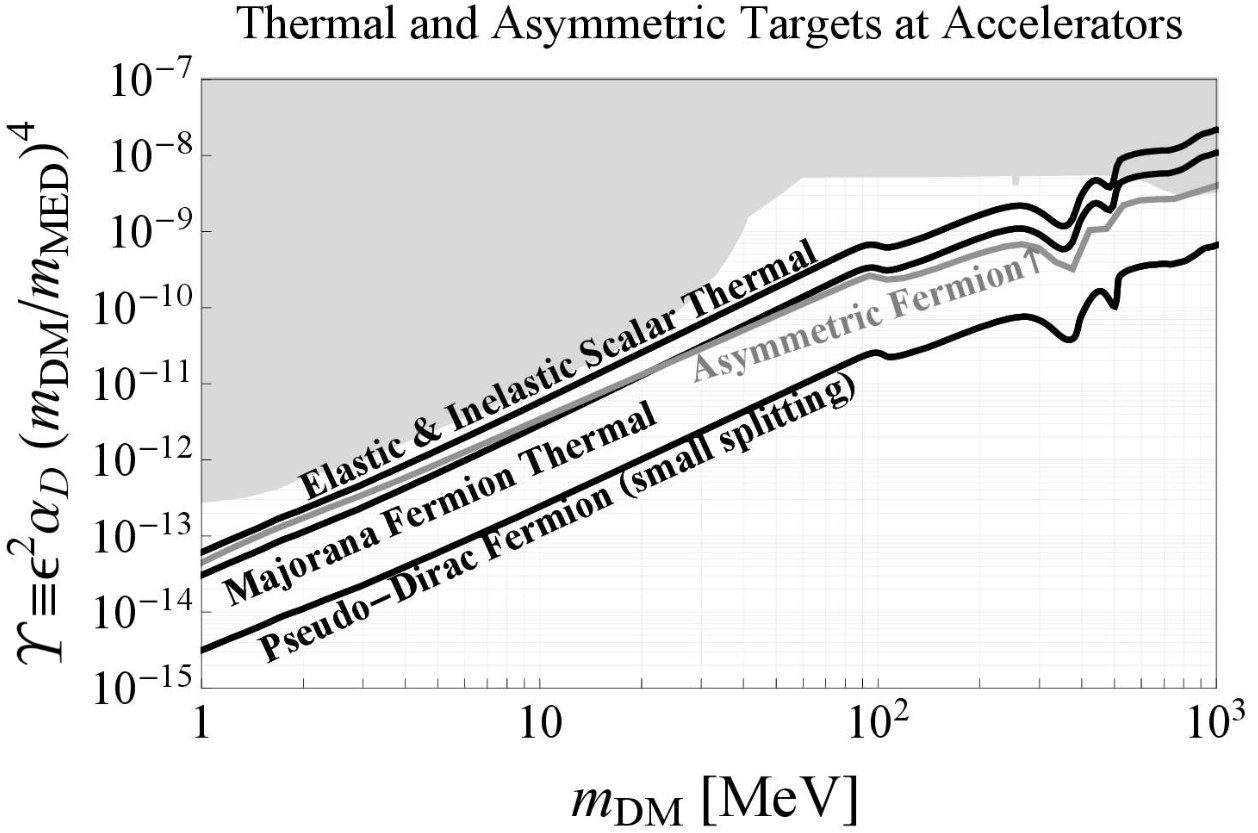}
    \caption{Thermal freeze-out targets (from direct annihilation) and asymmetric fermion DM target. This figure assumes the dark photon vector mediator introduced in the text with a mass ratio of $m_\dm / m_\med = m_\dm / m_\ap = 1/3$ and a dark fine structure constant of $\alpha_\dark = 0.5$. The region shaded in gray summarizes the current constrains. The plot was adapted from \cite{Battaglieri:2017aum}.}
    \label{fig:thermal_relics_vector}
\end{figure}

The simplest models giving rise to the DM current $J_\dark^\mu$ contain fundamental states with spin $s = 0$ or $1/2$.
Important benchmark models are:
Elastic scalar, inelastic scalar, Pseudo-Dirac, Majorana DM~\cite{Akesson:2018vlm} and asymmetric fermion DM~\cite{Kaplan:2009ag}, each of which come with a different thermal relic target in their parameter space.
For the dark photon as vector mediator, \cref{fig:thermal_relics_vector} (adapted from~\cite{Battaglieri:2017aum}) shows the thermal targets for the five benchmark DS models.
The plot makes conservative choices of $\alpha_\dark$ and $m_\dm / m_\ap$ and keeps $\varepsilon$ free.
The dark fine structure constant is assumed to be large, but still in the perturbative regime, $\alpha_\dark = 0.5$, and the mass ratio is set to be $m_\dm / m_\ap = 1/3$.
This benchmark is conservative in a sense that smaller values of $\alpha_\dark$ and $m_\dm / m_\ap$ \emph{improve} the experimental sensitivity, essentially because larger values of $\varepsilon$ are necessary to produce the correct relic abundance.
\\

The interaction term in \cref{eq:lepton_dp_interaction} leads to unique new physics effects potentially detectable in accelerator experiments.
The simplest scenario is the production of on-shell dark photons, which directly probe the parameter space spanned by $\{ m_\ap, \varepsilon\}$, i.e. the $\ap$ mass eigenvalue and the reduced kinetic mixing parameter.

\subsection{Theoretical Foundation of the \lohengrin Experiment}\label{sec:theory:dpatelsa}
In this section, we discuss the most dominant processes contributing to signal and background as well as the resulting demands on the detector layout.
The theoretical predictions presented in this section were obtained with a dedicated Monte Carlo code called \texttt{Lohengrin++}.
The squared amplitudes of the quantum mechanical processes were calculated using \texttt{FeynRules}~\cite{Alloul_2014}, \texttt{FeynArts}~\cite{Hahn_2001} and \texttt{FeynCalc}~\cite{MERTIG1991345}, while the Monte Carlo integration over the phase space is handled by the \texttt{Vegas} algorithm provided in the \texttt{CUBA} library \cite{Hahn:2004fe}.
For some final state particle $f$ we introduce a shorthand notation for lab frame phase space volumes associated to it, namely $\mathcal{Q}^f$, which will come with appropriate subscripts indicating applied cuts.

\subsubsection{Fundamental Signal Process}\label{sec:theory:dpatelsa:fundamental_process}
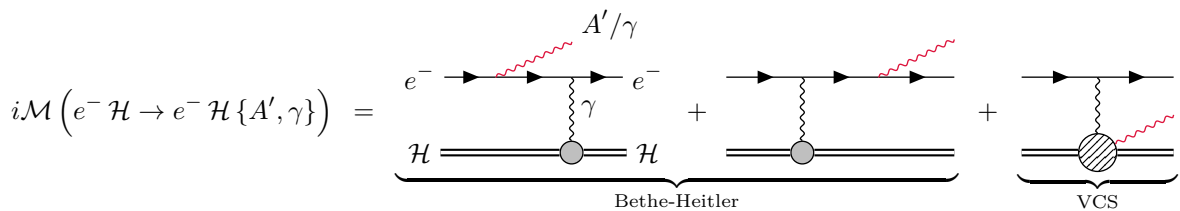
\begin{figure*}[htpb]
    \centering
    \begin{minipage}{\textwidth}
\begin{equation*}
    i \mathcal{M}\left( e^-\,\mathcal{H} \rightarrow e^- \, \mathcal{H} \, \{ A^\prime, \gamma \}\right) ~ = ~
        \underbrace{
        \begin{tikzpicture}[baseline=12.5]
        \setlength{\feynhandblobsize}{3mm}
        \setlength{\feynhandarrowsize}{5pt}
        \begin{feynhand}
        \vertex (ei) at (-1.5, 1) {$e^-$};
        \vertex (eo) at (1.5, 1) {$e^-$};
        \vertex (veg) at (0.5,1);
        \vertex (vedp) at (-0.5, 1);
        \vertex (dpo) at (1, 1.7) {$A^\prime / \gamma$};
        \vertex [grayblob] (vhg) at (0.5,0) {};
        \vertex (hi) at (-1.5,0) {$\mathcal{H}$};
        \vertex (ho) at (1.5,0) {$\mathcal{H}$} ;
        \propag [fermion] (ei) to (vedp);
        \propag [fermion] (vedp) to (veg);
        \propag [fermion] (veg) to (eo);
        \propag [photon] (veg) to [edge label =$\gamma$] (vhg) ;
        \propag [photon, color = crimson] (vedp) to (dpo);
        \propag [double,double distance=0.3ex,thick] (hi) to (vhg);
        \propag [double,double distance=0.3ex,thick] (vhg) to (ho);
        \end{feynhand}
        \end{tikzpicture}
        ~+~
        \begin{tikzpicture}[baseline=12.5]
        \setlength{\feynhandblobsize}{3mm}
        \setlength{\feynhandarrowsize}{5pt}
        \begin{feynhand}
        \vertex (ei) at (-1.5, 1);
        \vertex (eo) at (1.5, 1) ;
        \vertex (veg) at (-0.5,1);
        \vertex (vedp) at (0.5, 1);
        \vertex (dpo) at (1.5, 1.5);
        \vertex [grayblob] (vhg) at (-0.5,0) {};
        \vertex (hi) at (-1.5,0) ;
        \vertex (ho) at (1.5,0);
        \propag [fermion] (ei) to (veg);
        \propag [fermion] (veg) to (vedp);
        \propag [fermion] (vedp) to (eo);
        \propag [photon] (veg) to (vhg);
        \propag [photon, color = crimson] (vedp) to (dpo);
        \propag [double,double distance=0.3ex,thick] (hi) to (vhg);
        \propag [double,double distance=0.3ex,thick] (vhg) to (ho);
        \end{feynhand}
        \end{tikzpicture}
        }
        _{\text{Bethe-Heitler}}
        ~+~
        \underbrace{
        \begin{tikzpicture}[baseline=12.5]
        \setlength{\feynhandblobsize}{5mm}
        \setlength{\feynhandarrowsize}{5pt}
        \begin{feynhand}
        \vertex (ei) at (-1, 1) ;
        \vertex (eo) at (1, 1) ;
        \vertex (veg) at (0,1);
        \vertex (dpo) at (1, 0.5) ;
        \vertex [NEblob] (vhg) at (0,0) {};
        \vertex (hi) at (-1,0) ;
        \vertex (ho) at (1,0)  ;
        \propag [fermion] (ei) to (veg);
        \propag [fermion] (veg) to (eo);
        \propag [photon] (veg) to (vhg);
        \propag [photon, color = crimson] (vhg) to (dpo);
        \propag [double,double distance=0.3ex,thick] (hi) to (vhg);
        \propag [double,double distance=0.3ex,thick] (vhg) to (ho);
        \end{feynhand}
        \end{tikzpicture}
        }
        _{\text{VCS}}
\end{equation*}
\end{minipage}
\caption{Feynman diagrams contributing to the lowest-order amplitude of (dark) photon production in collisions of an electron with an hadronic system (nucleus or quasi-free nucleons). The red line indicates radiation of an on-shell (dark) photon. The grey blob represents form factor evaluations while the hashed blob represents the non-trivial VCS amplitude, which in general involves structure dependent parameters.}
\label{fig:Feynman_diagrams}
\end{figure*}

\lohengrin will search for the \emph{dark bremsstrahlung} process
\begin{eqnarray*}
 e^- + \mathcal{H} \rightarrow e^- + A^\prime + \mathcal{H},
\end{eqnarray*}
where an electron of energy $E$ scatters off a fixed-target hadronic system $\mathcal{H}$ to produce dark photons $A^\prime$ through the interaction term from \cref{eq:lepton_dp_interaction}.
The lowest order Feynman diagrams contributing to this process are displayed in \cref{fig:Feynman_diagrams}.

The dominant contribution arises from Bethe-Heitler (BH) scattering with one-photon exchange between the incoming electron and the nucleus, where the latter is characterized by its charge distribution.
To a good approximation, the nucleus can be treated as a scalar, since contributions arising from its magnetic moment are suppressed by its inverse mass.
Furthermore, a suitable target material, Tungsten, is naturally most abundant as three isotopes with $J^P = 0^+ (\ce{^{182}_{74}W}$ (27\% )$, \ce{^{184}_{74}W}$ (31\%)$, \ce{^{186}_{74}W}$ (28\%)) and a smaller fraction of a single isotope with $J^P = \frac{1}{2}^-$ (\ce{^{183}_{74}W} (14\%)).
The non-zero spin of the fermionic component will contribute to the cross section with terms scaling with the square of the inverse nuclear mass.
Hence they are strongly suppressed, and we will assume all target nuclei to be of scalar nature.
We denote the 
nucleus by the symbol $\varPhi^+$ and henceforth always work with the tungsten isotope $\varPhi^+ = \ce{^{184}_{74}W}$ which has a mass of $m_{\varPhi^+} \approx \SI{171.16}{\giga\electronvolt}$.
Due to its scalar nature, the target is characterized by only one form factor, for which we use an analytic approximation of the Woods-Saxon form factor. Its form can be found in \cite{Ballett:2018uuc}.

Note that in general, the (dark) bremsstrahlung process will also receive contributions from virtual Compton scattering (VCS), where the radiation is emitted from the composite hadronic system.
This is possible since the dark photon fundamentally also couples to quarks, see \cref{eq:lepton_dp_interaction}.
Moreover, the hadronic system can be generalized to not only account for the nuclear charge distribution of the target nucleus, but to also account for additional effects arising from the nuclear constituents, protons and neutrons, which also contribute via their charge and magnetic moments~\cite{Ballett:2018uuc}.
We found that neither the VCS nor effects from individual nucleons yield relevant contributions to the missing momentum search strategy.
The details are left for a dedicated theory focused publication.
The (experimental design driving) signal process is very well approximated by BH scattering off the target nucleus.

As mentioned earlier, \lohengrin is a missing-momentum experiment:
The search strategy relies on measuring the outgoing electron after dark photon emission, not on a reappearance of the decay signature of the dark photon.
Typically, the electron will receive a sizable transverse kick whose size is roughly controlled by the dark photon mass.

Since the fundamental production process is $2 \rightarrow 3$, the electron kinematics are not restricted to the $2 \to 2$ elastic line defined by energy-momentum conservation, but have three degrees of freedom, which we choose in standard spherical coordinates as:
Outgoing energy fraction $\xi = \frac{E_{\textrm{e,out}}}{E_{\textrm{e,in}}}$, scattering angle $\theta_e$ relative to the beam axis and azimuthal angle $\phi_e$.
The fundamental process is symmetric around the beam axis, such that the signal characteristics are fully described by a double differential cross section w.r.t. the energy fraction $\xi$ and electron solid angle $\varOmega_e$.

\begin{figure*}
    \centering
    \includegraphics[width=\textwidth]{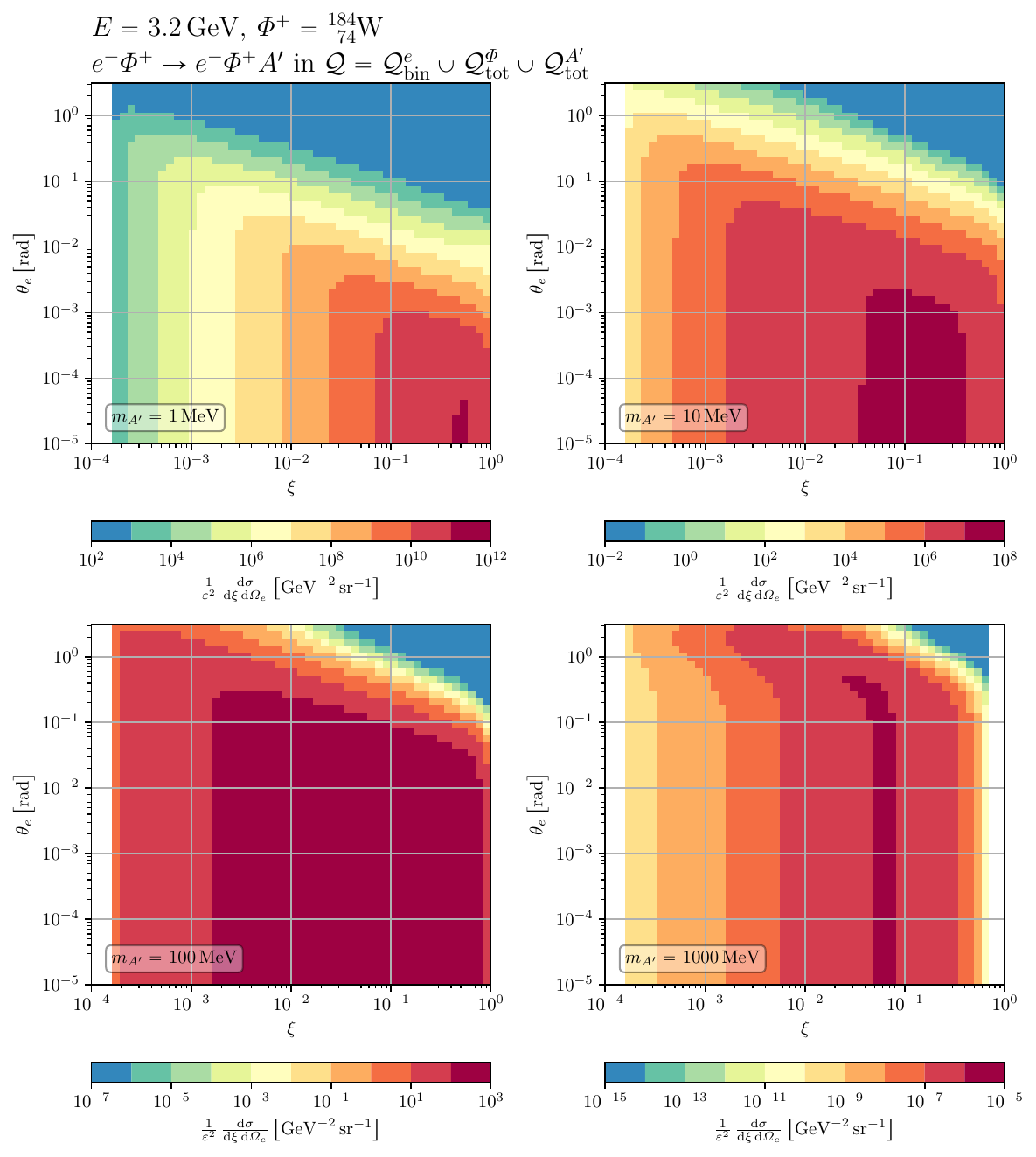}
    \caption{Double differential cross section of dark photon production w.r.t.\ the energy fraction $\xi$ and solid angle $\varOmega_e$ of the recoiling electron. The result is fully inclusive. We picked four different benchmark masses and normalized by $\eps^2$.}\label{fig:ddcs_sph_el_2to3DP_coherent_BH_4masses_ELSA}
\end{figure*}

The double differential signal cross section is shown in \cref{fig:ddcs_sph_el_2to3DP_coherent_BH_4masses_ELSA} for four benchmark dark photon masses and is normalized to $\varepsilon^2$.
For increasing dark photon mass, the electron kinematics shift towards lower energies and wider angles, which quantifies the transverse kick mentioned earlier.
The kinematically allowed final state electrons are located in a phase space window enclosed by a low-energy boundary $\xi \geq m_e / E$ and a high-energy boundary which depends nontrivially on the dark photon mass.
To a reasonable approximation (for $m_\ap \ll E$) this boundary can simply be thought of the elastic $2\rightarrow 2$ line.

\begin{figure}
    \centering
    \includegraphics[width=\columnwidth]{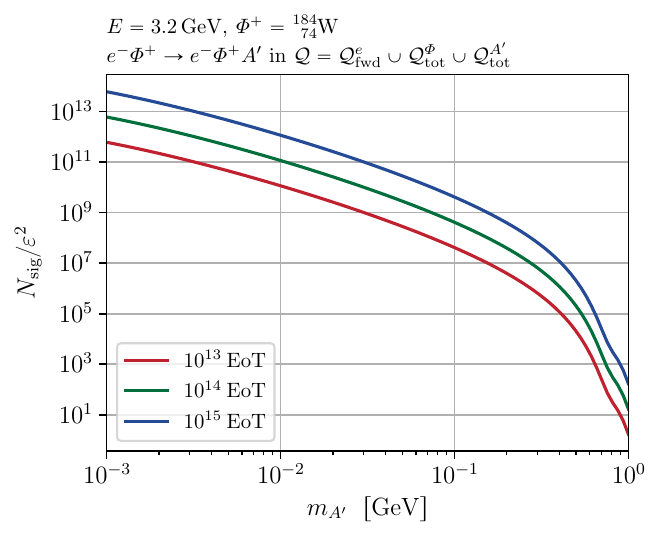}
    \caption{Number of signal events dependent on the dark photon mass, for different numbers of electrons on target. The event yield is normalized by the suppressing  parameter $\eps^2$ and takes only final state electrons in forward direction into account.}
    \label{fig:cs_fwd_ELSA_mDP}
\end{figure}

Assuming a tungsten target with a thickness of $0.1 X_0 \approx 0.35\,\textrm{mm}$, the expected yield of events with forward electrons ($\theta_e < \frac{\pi}{2}$) is shown in \cref{fig:cs_fwd_ELSA_mDP}, as a function of the dark photon mass for different numbers of electrons on target. For dark photon masses between $\SI{10}{\mega\electronvolt}$ and $\SI{100}{\mega\electronvolt}$, and SM to DS couplings at the relic targets, between 1 and 100 events with a dark photon radiated off the electron would be expected for $4\cdot10^{14}$ electrons on target. 

\subsubsection{QED Background Processes}\label{sec:theory:dpatelsa:fundamental_process_qed}
There are various processes that can mimic the signal signature. Due to the design of the \lohengrin experiment, SM bremsstrahlung events with a high energy photon that escapes detection are expected to be the dominant background source. This process is discussed in detail in this section; other backgrounds are discussed in \cref{sec:lohengrin:backgrounds}.

In QED, the lowest order process at $\mathcal{O}(\alpha^2)$ is $e^- + \varPhi^+ \rightarrow e^- + \varPhi^+$ that populates the elastic line set by energy-momentum conservation.
The large nucleus mass forces the elastic line, and so the outgoing electrons, to the region of high energy final states, $\xi \sim 1$.
This $2 \rightarrow 2$ process will thus not be of relevance when searching for missing momentum of $\mathcal{O}(\SI{}{\giga\electronvolt})$.

The leading background process is thus QED bremsstrahlung of $\mathcal{O}(\alpha^3)$,
\begin{eqnarray*}
    e^- + \varPhi^+ \rightarrow e^- + \gamma + \varPhi^+ ,
\end{eqnarray*}
where the radiated photon carries away a large fraction of the incoming electrons energy (i.e. $\xi \ll 1$) and escapes detection.
The corresponding Feynman diagrams are displayed in~\cref{fig:Feynman_diagrams}.
Once more, it is sufficient to restrict ourselves to Bethe-Heitler scattering off the scalar nucleus.

The emission of hard photon radiation for this process is dominated by configurations where the photon is emitted collinearly with respect to the incoming/outgoing electron direction.
Additional configurations also exist in which a relatively high-energy photon is emitted at wide-angle, leaving behind a sufficiently low energy electron without accompanying photon radiation within its angular vicinity.

\begin{figure}[t]
    \centering
    \includegraphics[width=\columnwidth]{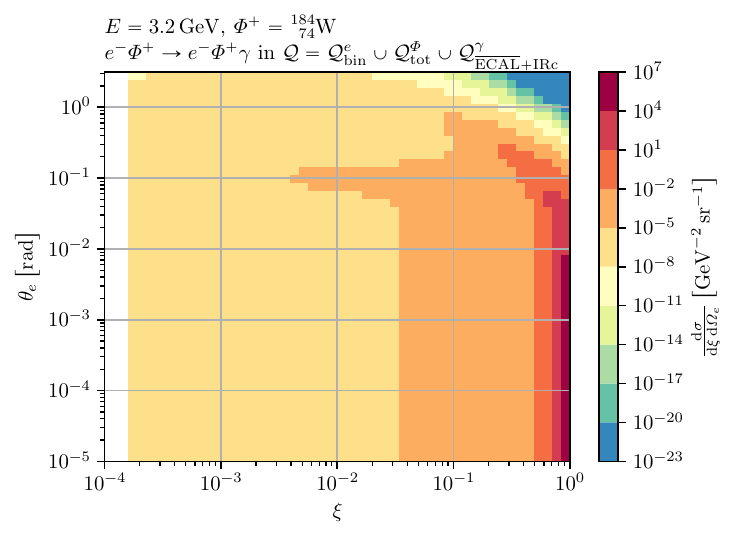}
    \caption{Double differential cross section w.r.t. the energy fraction $\xi$ and solid angle $\varOmega_e$ of the recoiling electron for QED bremsstrahlung. In this result, only events with infrared safe photons that miss the ECAL are taken into account.  }\label{fig:ddcs_sph_el_2to3SM_coherent_full+thgammacut_xcut10MeV_BH}
\end{figure}

The above process thus requires a reliable veto of the QED photon with an electromagnetic calorimeter (ECAL). The coverage of the ECAL will be limited to the forward direction due to the overall design of the \lohengrin experiment: 
hard photons scattered to wide angles miss the forward ECAL, and cannot be vetoed. This introduces an irreducible, but well understood background.
In \cref{fig:ddcs_sph_el_2to3SM_coherent_full+thgammacut_xcut10MeV_BH} we show the double differential cross section for the \emph{irreducible} background, again in terms of the observable electron kinetic variables.

\begin{figure}
    \centering
    \includegraphics[width=\columnwidth]{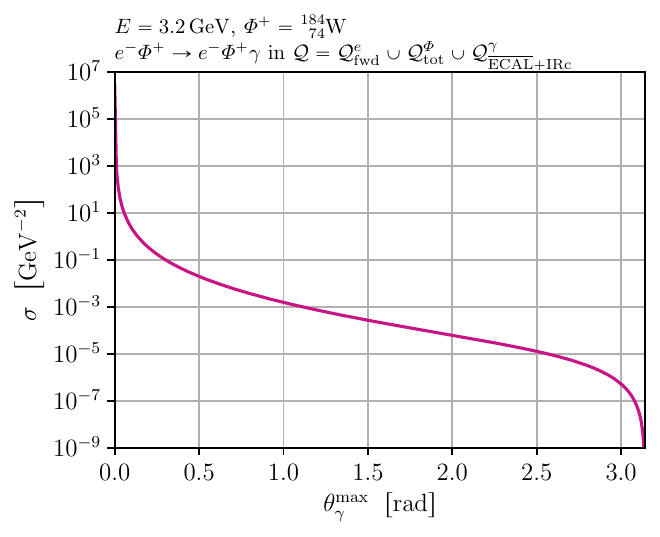}
    \caption{Cross section of the leading QED background, dependent on the maximally veto-able photon angle (which corresponds to the lower integration limit).}
    \label{fig:cs_th_gamma_cut}
\end{figure}

\begin{figure*}
    \centering
    \includegraphics[width=\textwidth]{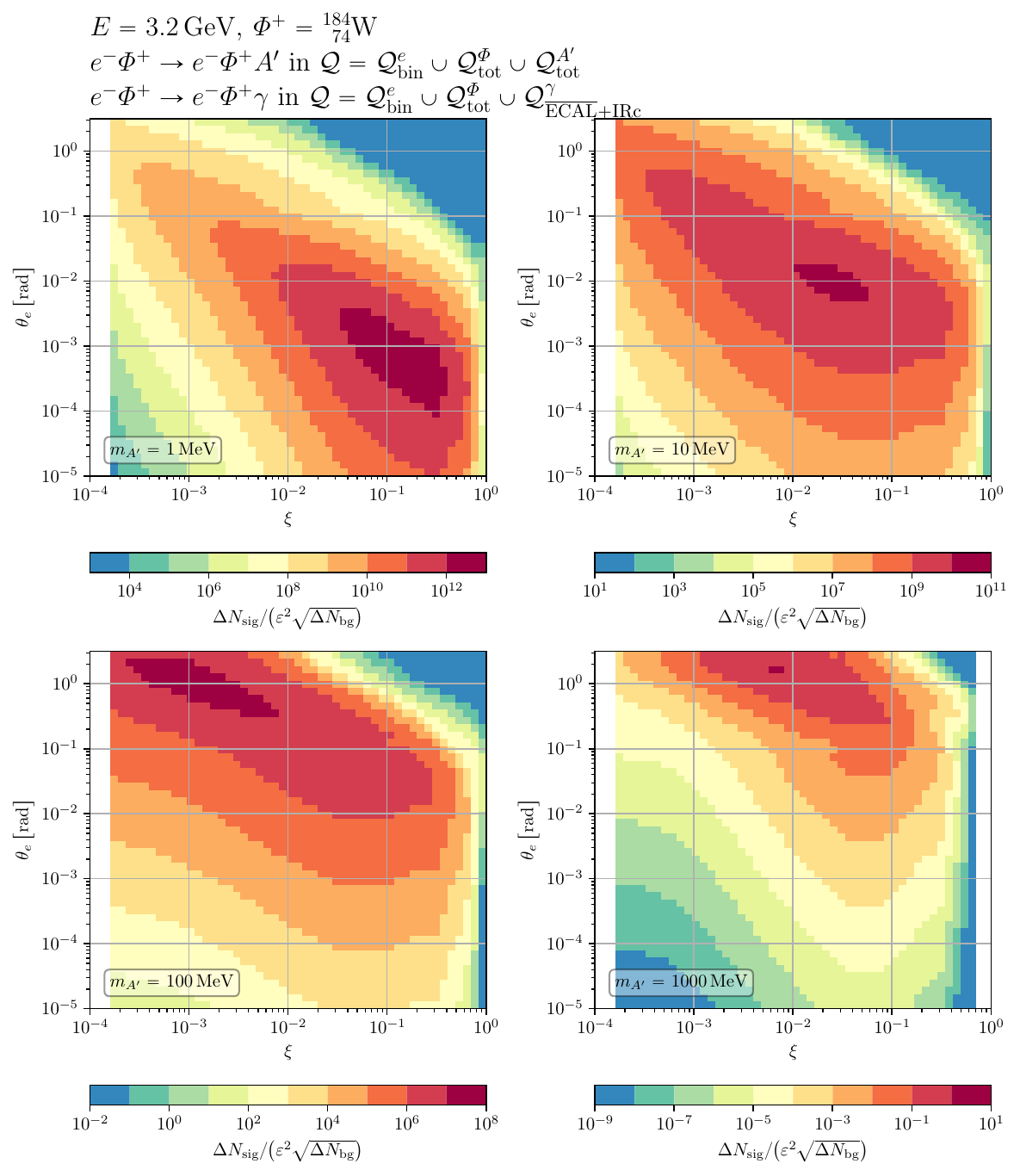}
    \caption{Expected sensitivity $\frac{S}{\sqrt{B}}$ for $\varepsilon = 1$ for four different dark photon masses as a function of the recoil electron energy and scattering angle.}
    \label{fig:SoverSqrtBG_sph_el_2to3DP_coherent_BH_4masses_ELSA}
\end{figure*}

Here we assume a perfect ECAL veto in a forward cone set by a polar angle of size $\SI{0.1}{\radian}$, meaning that the irreducible background consists of photons scattered to angles $\theta_\gamma > \SI{0.1}{\radian}$.\footnote{We also introduced a cutoff of low energy photons to regulate the infrared divergence.
Such a cutoff is naturally provided by the ECAL resolution, which we assume to be $\sim\SI{10}{\mega\electronvolt}$. For low energy electrons below $\sim\mathcal{O}(\SI{}{\giga\electronvolt})$, the background is not sensitive to the exact value of this cutoff.} \cref{fig:cs_th_gamma_cut} shows the cross section of irreducible background events with forward electrons ($\theta_e > \frac{\pi}{2}$) as a function of the maximum angle covered by the ECAL, $\theta_\gamma^\text{max}$ (i.e. integrated over the interval $[\theta_\gamma^\text{max}, \pi]$ of photon angles). The cross section obviously peaks for forward photons, but clearly a larger ECAL coverage greatly benefits the overall sensitivity of the experiment by reducing the SM bremsstrahlung background exponentially.
One objective of experimental optimization is thus to maximize the solid angle coverage of the ECAL. The expected sensitivity is estimated here as the quantity $S/\sqrt{B}$, where $S$ is the expected number of signal events and $B$ is the expected number of background events. For a setup with a perfectly efficient tracking detector for electrons, and a perfectly efficient ECAL for photons covering an opening angle of $\theta_{\gamma,\text{max}} = 0.1\,$rad, the expected sensitivity as a function of the electron kinematics is shown in \cref{fig:SoverSqrtBG_sph_el_2to3DP_coherent_BH_4masses_ELSA}. For dark photon masses between \SI{1}{\mega\electronvolt} and \SI{100}{\mega\electronvolt}, and neglecting additional background sources that are addressed below, a signal region containing a final state electron with an energy between \SI{1}{\percent} and \SI{10}{\percent} of the incoming electron's energy that is scattered at angles below \SI{0.1}{\radian} seems most promising. 

%% file: lohengrinexperiment.tex
The analysis presented in section \ref{sec:theory:dpatelsa} sets the goalposts for the \lohengrin experiment: a detector is required that can deal with an event rate of \SI{100}{\mega\hertz}. The detector must provide good tracking efficiency and resolution for low momentum electrons with a momentum as low as few tens of MeV that emerge from the target at moderate scattering angles. In addition, a reliable veto system is required for SM photons that are radiated off the electrons in the target. Furthermore, additional veto systems must be employed for other SM backgrounds - these comprise events with neutral hadrons that are produced in electro-nuclear or photo-nuclear interactions in the target and/or the detectors.

Considering these goalposts, the sensitivity of the \lohengrin experiment to the production of dark bremsstrahlung stems from six crucial building blocks: (1) the ELSA accelerator allows for the extraction of single electrons with an energy of \SI{3.2}{\giga\electronvolt} at a high rate that can be directed onto a thin target; the target is placed in (2) a strong magnetic field that is sufficient to bend the trajectories of electrons with an energy of up to \SI{3.2}{\giga\electronvolt} around the calorimeter systems; (3) a tracker consisting of several layers of ultrathin and ultrafast silicon pixel detectors is placed around the target in order to tag incoming electrons and reconstruct the tracks of scattered electrons down to the lowest energies behind the target; (4) an ECAL is placed behind the tracker in order to measure the energy of SM bremsstrahlung photons in the final state; the fact that the magnet diverts most of the scattered electrons away from the ECAL enables the efficient identification of rare events with high energy photons being emitted in the target; the ECAL is embedded in (5) a coarse hadronic calorimeter (HCAL) that acts as a veto system for any hadronic energy in the final state. In order to maintain a high rate of incoming electrons (6) a two stage track trigger system is used to efficiently select events with low energy electrons in the final state, rejecting any events with high energy electrons behind the target.

\subsection{The ELSA Accelerator and Beamline}
\begin{figure}
    \centering
    \includegraphics[width=\columnwidth]{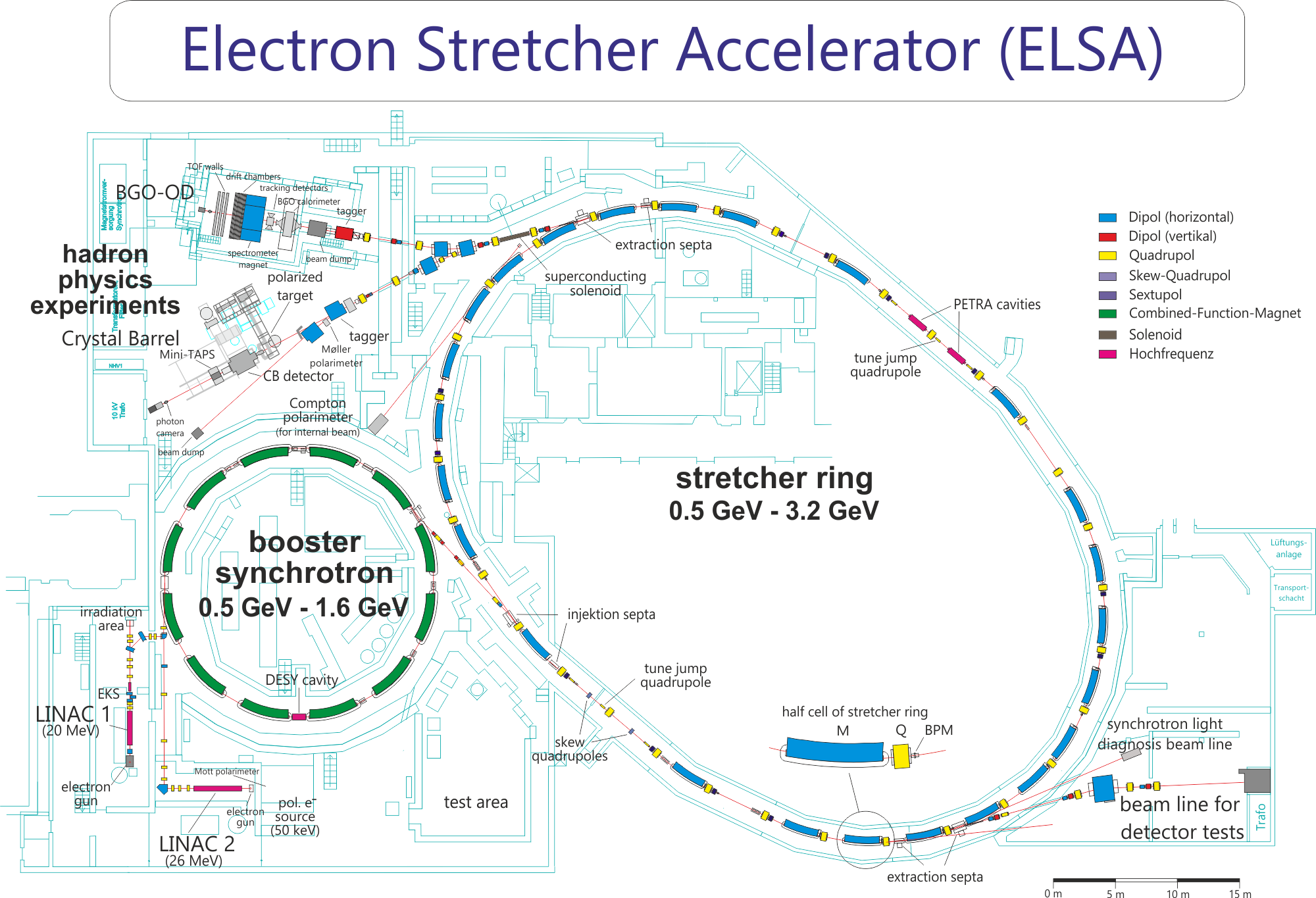}
    \caption{Schematic view of the ELSA accelerator in its experimental hall \cite{ELSAHomepage}.}
    \label{fig:ELSAOverview}
\end{figure}

The Electron Stretcher Accelerator (ELSA) is an electron accelerator located at the ``Physikalisches Intitut" of the University of Bonn. A general overview of the experimental hall with the accelerator situated inside can be found in \cref{fig:ELSAOverview}.

It is the ELSA accelerator in particular that enables the full potential of the \lohengrin experiment, as - on average - a single electron per event can be extracted from the storage ring, providing a very clean and manageable initial state. The small energy spread of the initial state electrons of \SI{0.8}{\permille} allows to rely on a rather basic tagging detector, minimizing the amount of material in front of the target. In combination with the small number of incident electrons per event the initial state is hence precisely known. 

Based on a previous analysis of the optics for the detector test beamline at ELSA~\cite{handle:20.500.11811/7310}, a reasonable approximation for the beamspot on the target is a gaussian profile with a standard deviation of \SI{1}{\milli\meter} in both lateral dimensions, with a divergence of less than \SI{0.8}{\milli\radian} in both dimensions. These are the properties that have been used for the present analysis. Whether a circular beam spot on the target is preferable over a more stretched or compressed beam spot will be explored in the future, as will the impact of the beam divergence on target. 

With a bunch spacing of \SI{2}{\nano\second} in the accelerator, an average of \num{0.2} electrons will be extracted from each bunch in each revolution, yielding an average extraction rate of \SI{100}{\mega\hertz}. The probability to extract more than one electron in a single event is less than \SI{1.8}{\percent} for this configuration.

\subsection{Detector}\label{sec:lohengrin:detector}
\begin{figure}
    \centering
    \includegraphics[width=\columnwidth]{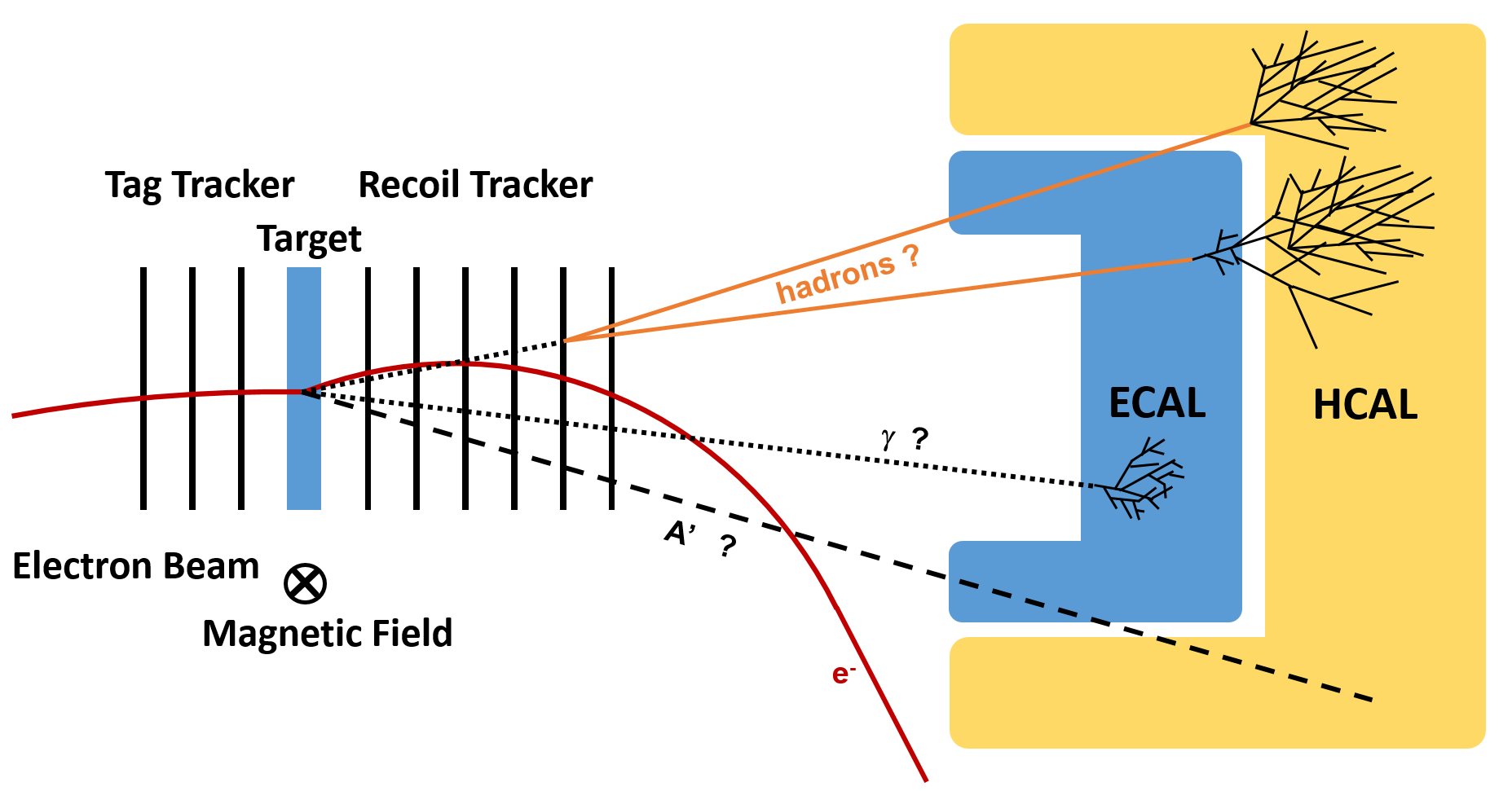}
    \caption{Schematic drawing of the detector setup for \lohengrin. It consists of the tagging tracker, target, recoil tracker, electromagnetic calorimeter, hadronic calorimeter and a magnetic field.}
    \label{fig:LohengrinSchematic}
\end{figure}

\begin{figure}
    \centering
    \begin{tikzpicture}
        \node[anchor=south west,inner sep=0] (Bild) at (0,0) {\includegraphics[width=\columnwidth]{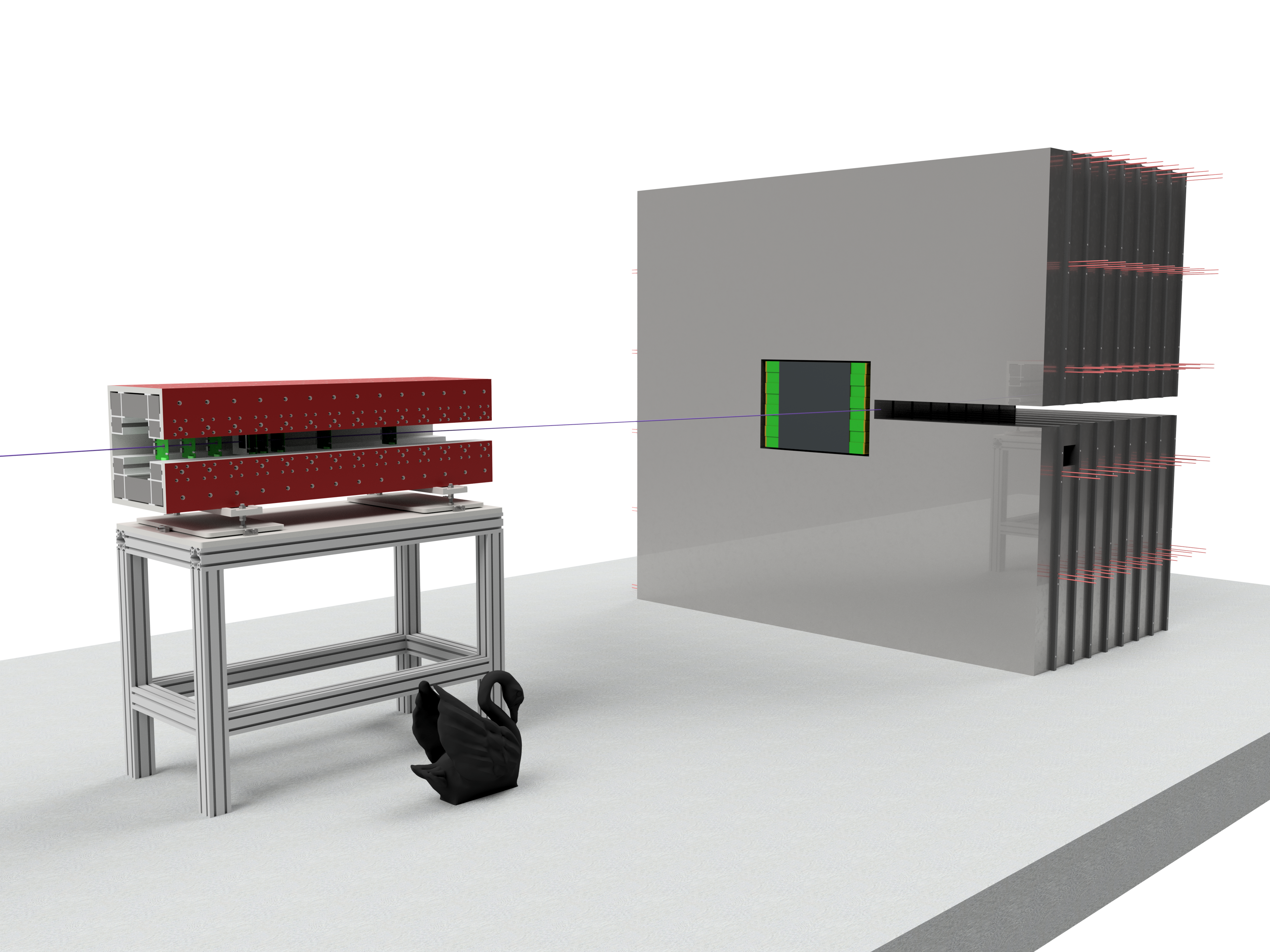}};
        \begin{scope}[x=(Bild.south east),y=(Bild.north west)]
            \node [right] (magnet) at (0.47,0.275){Magnet};
			\draw [-stealth, thick, black](magnet.north) -- (0.39,0.51);
			\node [right] (tagg) at (0,0.775){Tagging tracker};
			\draw [-stealth, thick, black](tagg.south) -- (0.15,0.525);
			\node [right] (target) at (0.18,0.7){Target};
			\draw [-stealth, thick, black](target.south) -- (0.195,0.53);
			\node [right] (rec) at (0.26,0.85){Recoil Tracker};
			\draw [-stealth, thick, black](rec.south) -- (0.25,0.535);
			\node [left] (ecal) at (0.85,0.2){ECAL};
			\draw [-stealth, thick, black](ecal.north) -- (0.65,0.55);
			\node [right] (hcal) at (0.6,0.9){HCAL};
			\draw [-stealth, thick, black](hcal.south) -- (0.7,0.8);
			\node [right] (electron) at (0,0.56){$\text{e}^-$};
        \end{scope}
    \end{tikzpicture}
    \caption{CAD rendering of the detector setup for \lohengrin. It consists of the tagging tracker, target and the recoil tracker inside the magnet and also the electromagnetic calorimeter (ECAL), hadronic calorimeter (HCAL) towards the rear. The path of the non interacting electrons is indicated in purple. Cygnus atratus is added for size comparison.}
    \label{fig:LohengrinOverviewRendering}
\end{figure}
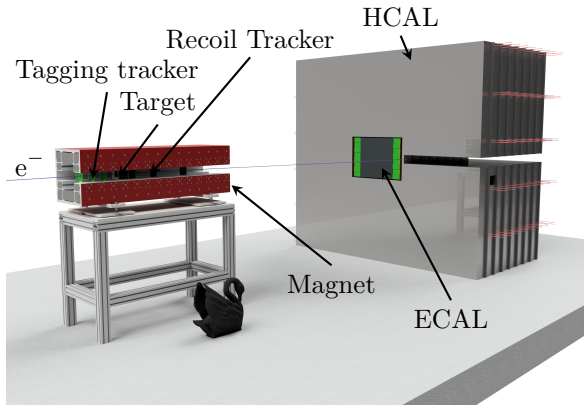

A schematic overview of the \lohengrin experiment is shown in \cref{fig:LohengrinSchematic}. In addition, \cref{fig:LohengrinOverviewRendering} shows a three-dimensional rendering of the proposed detector assembly. The \lohengrin tracker and target are situated within a permanent magnet that bends the trajectories of the incident electrons such that they hit the target perpendicularly on average. The same magnet allows for a precise track reconstruction and momentum measurement for electrons in the final state. The ECAL covers a solid angle that covers a significant fraction of the opening angle of the magnet in the forward direction, allowing a precise measurement of final state photons. The HCAL surrounds the ECAL, covering a larger solid angle. An extension of the HCAL towards and possibly around the magnet will be discussed below.

A key feature of the proposed experiment is the combination of the bending power of the magnet and the location and coverage of the ECAL. For dark photon masses between \SI{10}{\mega\electronvolt} and \SI{100}{\mega\electronvolt}, the sensitivity of the experiment must extend to values between $10^{-6}\lesssim \varepsilon \lesssim 10^{-4}$, roughly, in order to reach the relic target. As can be read from \cref{fig:SoverSqrtBG_sph_el_2to3DP_coherent_BH_4masses_ELSA}, this requires an amount of at least $10^{14} - 10^{15}$ electrons on target. 
Hence we choose an average rate of \SI{100}{\mega\hertz} of electrons on target as a baseline. This allows the conclusion of the experiment within one year after construction (considering the duty cycle of ELSA, the limited available beamtime, etc.). While ELSA can deliver single electrons at a higher rate, this baseline rate already poses a considerable challenge for the readout and timing of all sub-detectors. A much lower rate would increase the projected duration of the experiment beyond reasonable limits. 

While challenging, tracking single electrons at a rate of \SI{100}{\mega\hertz} seems feasible, considering the recent advances in the design of depleted monolithic active pixel sensors~\cite{PERIC2007876, Wermes_2015}. The precise measurement of the electron energy in the final state through the ECAL is significantly more difficult in the proposed setup, however. This is due to the small size of the beamspot on target - most electrons will traverse the target loosing very little energy and emerge from the target at a small polar angle. As the granularity of the ECAL is significantly coarser than the granularity of the tracking detector, most final state electrons would hit the same ECAL cells, making an event based energy measurement extremely challenging. A different approach is therefore chosen for the \lohengrin experiment: the bending power of the magnet in combination with the location of the ECAL is chosen such that the trajectories of most final state electrons, regardless of their total momentum, are bent around the calorimeters, and are only reconstructed from tracking data. The implications of this approach are discussed in more detail in \cref{sec:lohengrin:detector:ecal,sec:lohengrin:physics}. The ECAL hence only measures the energy of photons that are emitted below a certain polar angle in the target. This approach allows for the efficient identification of high energy photons over the background of low energy photons in the final state, while keeping the signal efficiency at a reasonable level. The expected rate of events with hadronic energy deposits is low (one event with hadronic energy deposits is expected for roughly one million electrons on target) and the analysis strategy aims to veto events with any hadronic activity in the final state - a precise measurement of the hadronic energy in the final state is hence not necessary, but in order to maintain a high sensitivity the HCAL must have a high efficiency for neutral hadrons at very low noise levels.

With dimensions of $\sim 4\times6\,\textrm{m}^2$, the proposed experiment will comfortably fit into the available space at the experimental areas at ELSA \mbox{($> 6\times8\,\textrm{m}^2$).}

\subsubsection{Target}\label{sec:lohengrin:detector:target}
The target is the one single component of the experiment which comprises most of the material budget. The layout of the experiment is hence optimised for dark photon and SM bremsstrahlung reactions occuring here. 

Material and thickness of the target must be carefully chosen. A thicker target increases the likelihood for dark photon production, but comes with the downsides associated with a larger material budget. This includes a more difficult reconstruction of recoiled electrons and an increase in the rate of high energy photons in the final state. We have chosen a thickness of $\num{0.1} X_0$ as this gives a reasonably thin material budget.

Tungsten as a target material has several benefits. The small radiation length allows for a physically thin target. 
Furthermore the main isotope of tungsten is a scalar nucleus. This simplifies signal modeling in a first approximation. Other materials for the target are investigated with respect to the number of background events with hadronic activity in the final state. For the present analysis, tungsten is used as a baseline material for the target.

\subsubsection{Magnet}\label{sec:lohengrin:detector:tracker}
At this stage, a permanent magnet providing a magnetic field
with a flux density of $B\approx \SI{1}{\tesla}$ is considered. 
Permanent magnets have the advantage of being much cheaper to set up and operate. Having this magnetic field strength over a length of $\sim$\SI{1}{\meter} provides enough bending power to steer the electron beam away from the electromagnetic calorimeter, as is required to reduce the necessary hit rate in the calorimeters to an acceptable level. In order to minimise the pollution of the tracking volume with backscattered electrons and secondaries that could be produced in the interaction of the primary electron beam with the magnet material, the proposed design features an opening along the side of deflection. An iron dominated magnet with such a ''C-shape'' seems a feasible option to provide the required bending power over a length of \SI{1}{\meter}, roughly. 

At this point we consider a simple dipole field pointing in the direction of the positive y-axis for the magnet as the baseline. A more sophisticated magnet design is currently being studied. This would allow an increase in the coverage of the calorimeters, possibly reducing the contamination of the signal region with background events.

\subsubsection{Tracking Detector}
\label{sec:lohengrin:detector:trackerDetailed}
The tracking detector is placed around the target inside of the bore of the magnet. The \lohengrin tracker will consist of a number of ultra-thin silicon pixel detectors, of which three layers are placed in front of the target in order to establish the presence of a beam electron in the initial state, and a number of layers behind the target to enable the tracking of scattered electrons in the final state. 

The requirements on the tracking detector are challenging: first, it must be able to deal with a high hit rate of $10^{8}$ electrons per second or more with a relatively narrow spatial distribution. Second, it must tag incoming electrons with an efficiency that is close to \SI{100}{\percent}, and it must be able to track scattered electrons over a large momentum range from \SI{25}{\mega\electronvolt} to \SI{3.2}{\giga\electronvolt}. For high energy electrons with a momentum of several hundred MeV or more the most important requirement is a high tracking efficiency, here the momentum resolution does not have to be extraordinarily good. It is much more challenging to achieve a satisfactory momentum resolution on the lower end of the targeted momentum spectrum. The material budget of the tracking detector must be minimised in order to maintain a high tracking efficiency with a reasonable momentum resolution in this part of the final state electron phase space. For this reason we consider depleted, active, monolithic pixel sensors as the only viable candidate\footnote{Ultrathin hybrid pixel detectors could also be an option; however, the development of suitable DMAPS is more advanced, and with the Monopix ASICs candidate designs exist that provide the required qualitative functionality while falling short of some of the quantitative requirements.}. We note that a new generation of these ASICs is required for a successful execution of the \lohengrin experiment: the ASICs must provide a configurable, fast hit-or signal that indicates whether a specific region of the ASIC was hit and a fast shaping of the analog signal. The required performance could possibly be achieved at the cost of limited or no time-over-threshold information, limiting the tracking resolution. 

A preliminary design of the tracking detector is based on the TJ-Monopix2 ASIC, a DMAPS that features square pixels with a pitch of \SI{33.04}{\micro\meter} in a matrix of $512\times512$ pixels~\cite{MOUSTAKAS2019604, doi:10.7566/JPSCP.42.011021}. Each tracking plane consists of 4 such ASICs, yielding a total size of $\sim 3.4\times3.4$\,cm$^2$ in the x-y-plane. The planes are centered approximately around $x=y=0$, at the z positions that are shown in \cref{tab:TrackerPositions}. A more optimised layout, that will also take into account the fact that the electrons' trajectrories are bent towards the positive x-axis, will be studied in the future to maximise the tracking efficiency.

Measurements with the TJ-Monopix2 ASICs demonstrate a single hit efficiency for electrons (measured at a low incident rate) at \SI{99.96\pm 0.04}{\percent}. The current analog FE of the TJ-Monopix2 ASIC is, however, not suitable for the \lohengrin experiment: the shaping time must be decreased significantly, which can be achieved at the cost of limited energy and timing resolution. Assuming a linear shape for the measured signals with a peaking time of \SI{5}{\nano\second} and a return to baseline within \SI{95}{\nano\second} after peaking, the estimated hit efficiency decreases to \SI{99.86\pm 0.05}{\percent} for the baseline beamspot on target. Implications of this choice are considered for the final sensitivity estimate in \cref{sec:lohengrin:physics}.

\subsubsection{Trigger System}\label{sec:lohengrin:detector:tdaq}
The \lohengrin trigger system must select events with low energy electrons. 
This has to be done with a high efficiency while reducing the overall event rate from \SI{100}{MHz} to about \SI{1}{MHz} of recorded events. This is achieved in a two stage process.

\begin{figure}[ht!]
\centering
\includegraphics[width=0.48\textwidth]{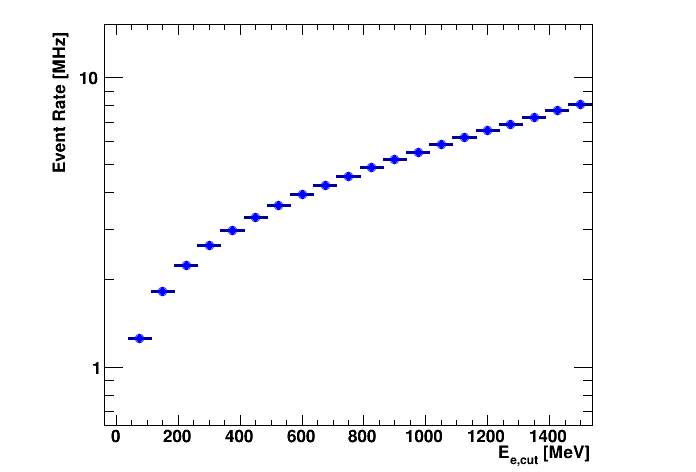}
\caption{Event rate vs the maximum energy of all electron(s) in the final state for SM bremsstrahlung events.}
\label{fig:LohengrinSetup:RateVsEnergy}
\end{figure}

In the first stage, an ultrafast hit-or signal from the readout ASICs in the tracking detector is used. This so-called L0 trigger exploits the fact that the trajectories of low energy electrons will be strongly bent in the magnetic field and will hit the first tracking planes far away from the central region, as opposed to high energy electrons that will hit tracking planes that are close to the target centrally. The goal for the first stage of the trigger system is to reduce the event rate by a factor of $\sim10-20$. The expected rate of SM bremsstrahlung events with any number of electrons in the final state, where the leading electron has an energy of at most $E_{e,\text{cut}}$ is shown in \cref{fig:LohengrinSetup:RateVsEnergy}. In order to achieve the targeted rate reduction, the L0 trigger must efficiently select events with a maximum electron energy of \SI{150}{MeV} or less, while efficiently vetoing events with electrons in the final state above this threshold. \\
A simple candidate trigger that achieves this goal is presented in the following; the estimated performance of this candidate system can be considered conservative, as many aspects can possibly be improved before deployment. 
Whenever a hit is registered in a configurable region of a given ASIC, the hit-or signal is set high, and is reset to low after a configurable time, e.g. 2\,ns\footnote{It is understood that this poses a considerable challenge on the tracking ASIC design, as the implementation of such a fast signal over the full pixel matrix will be difficult}. Assuming that such a signal can be implemented, the candidate L0 trigger relies on four tracking planes which are placed close to the target, at distances of 1\,cm, 3\,cm, 4.5\,cm and 7\,cm. If a hit is registered anywhere in the first plane, and at a value of x of at least 1.99\,mm/2.3\,mm/3.1\,mm in either of the second/third/fourth plane, the trigger fires. If that is not the case, or if a hit is registered in a fifth tracker plane that is placed at a distance of at least 10\,cm from the target, the event is discarded.

\begin{figure}[ht!]
\centering
\includegraphics[width=0.45\textwidth]{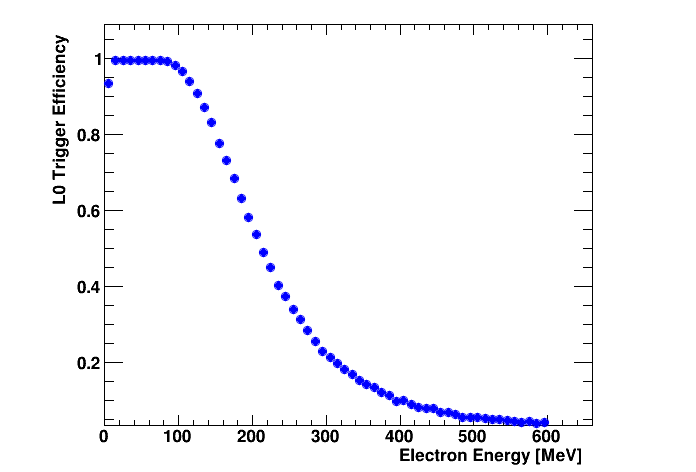}
\caption{Efficiency of the proposed L0 trigger as a function of the electron energy for $\theta_e < 0.05$.}
\label{fig:LohengrinSetup:TriggerEfficiencyVsE}
\end{figure}

\begin{figure}[ht!]
\centering
\includegraphics[width=0.45\textwidth]{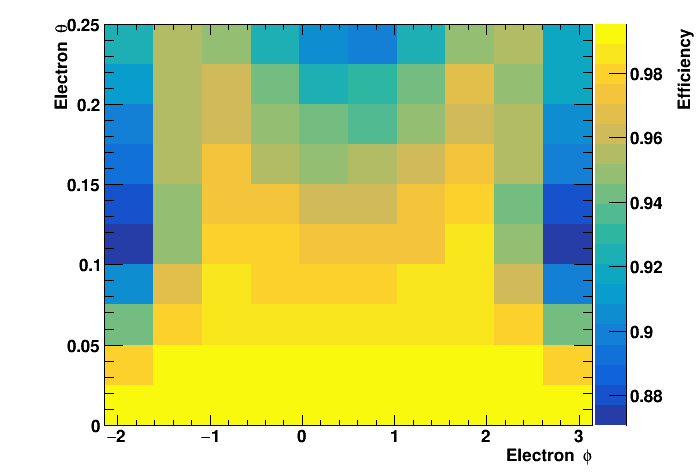}
\caption{Efficiency of the proposed L0 trigger as a function of the final state electron angles $\theta_e$ and $\phi_e$ for \mbox{$E_e < \SI{100}{\mega\electronvolt}$}.}
\label{fig:LohengrinSetup:TriggerEfficiencyVsThetaPhi}
\end{figure}

The efficiency of such a L0 trigger depends on the position of the electron hit in the target, as well as on the energy, the polar scattering angle and the azimuthal scattering angle of the electron in the target. For a beam with a Gaussian profile with a mean value of $0$ and a standard deviation of \SI{1}{\milli\meter} in both lateral dimensions, the expected efficiency of such a trigger, using a sample of SM bremsstrahlung events and a single hit efficiency of \SI{99.5}{\percent} is shown in \cref{fig:LohengrinSetup:TriggerEfficiencyVsE} as a function of the electron energy for $\theta_e < 0.05$ and as a function of $\theta_e$ and $\phi_e$ for $E_e < \SI{100}{\mega\electronvolt}$ in \cref{fig:LohengrinSetup:TriggerEfficiencyVsThetaPhi}. For electrons with an energy of less than \SI{100}{MeV}, an efficiency of approximately \SI{99}{\percent} is achieved, while the total event rate is reduced from \SI{100}{\mega\hertz} to about \SI{3.7}{\mega\hertz}. 

In order to reduce the amount of data which must be written to disk even further, a second trigger stage based on hardware AI engines is currently being studied. A graph neural network will be run on the AI engines for pattern recognition and coarse track fitting. This will enhance the resolution and reduce the fake efficiency of the L0 trigger. It is expected that the output rate of the L0 trigger is compatible with such an approach.

\subsubsection{Electromagnetic Calorimetry}\label{sec:lohengrin:detector:ecal} 
The electromagnetic calorimeter is primarily used to veto events that contain a high energy SM photon. It will be exposed to a high rate of low energy photons. However it must be able to accurately measure a large amount of energy deposited by a single high energy photon. This has to be done over a background of events selected by the low momentum electron trigger.
It must be sufficiently radiation hard, properly functioning after shooting $\sim 10^{15}$ electrons on target.

While crystal calorimeters provide good timing resolution, common candidates using CsI and BaF$_2$ crystals have been found to be limited by radiation hardness for \lohengrin, in particular in the central region where most of the bremsstrahlung photons are expected to hit the detector.

\begin{figure}
    \centering
    \includegraphics[width=\columnwidth]{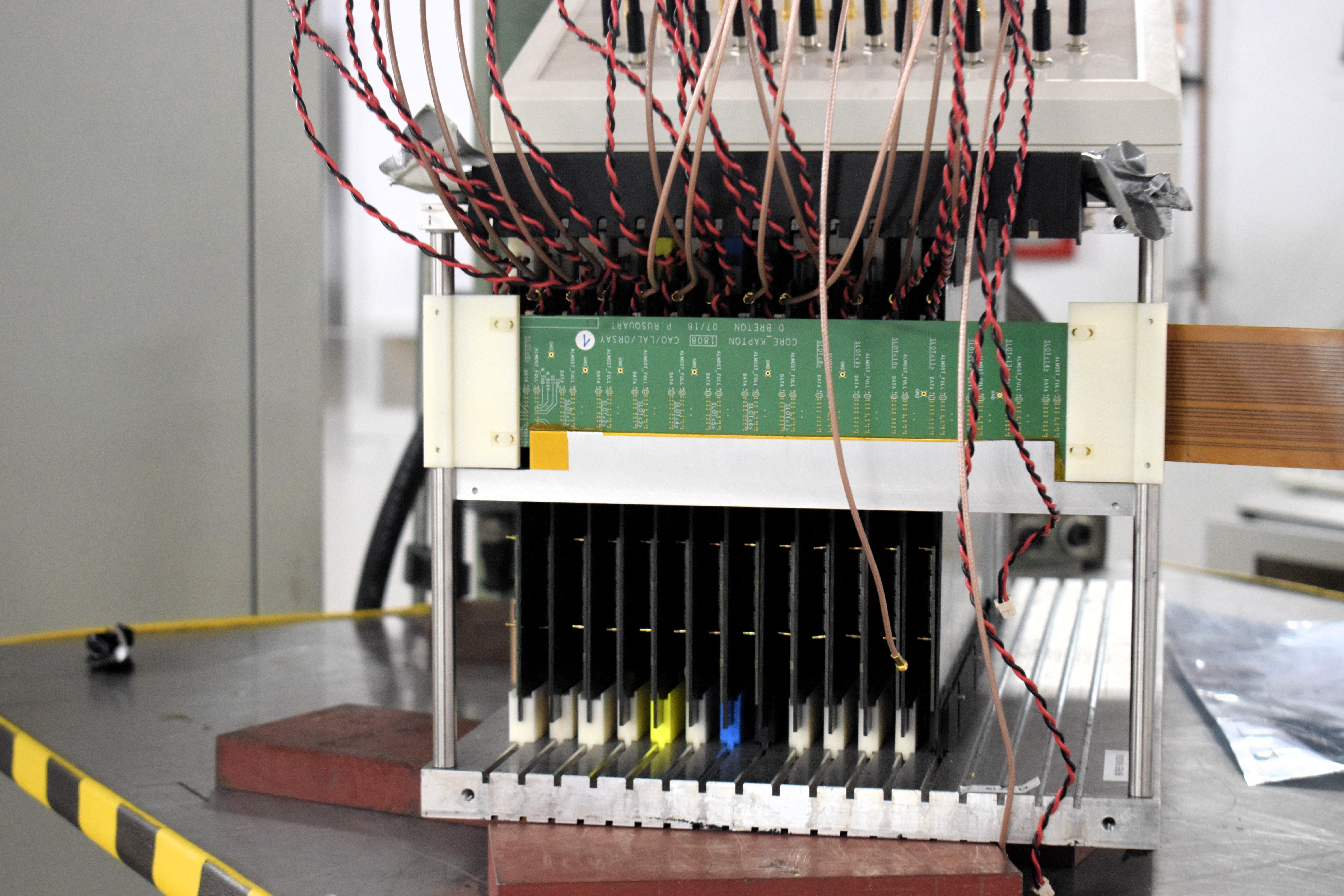}
    \caption{Photograph of several layers of the CALICE calorimeter prototype from 2021.}
    \label{fig:CaloPicture}
\end{figure}

As an alternative, a silicon tungsten sampling calorimeter has been studied and seems to be a feasible solution for the proposed experiment. For the preliminary studies, a SiW sampling calorimeter which is based on a prototype built by the CALICE Collaboration~\cite{PoeschlCalicePrototypeStatus} has been included in the simulation. It consists of 30 absorber layers made of \SI{2}{\milli\meter} thick tungsten plates followed by segmented silicon sensors and readout chips each. Each layer covers an active area of $\num{18}\times\SI{18}{\centi\meter\squared}$, while each pad has a size of $\num{5.5}\times\SI{5.5}{\milli\meter\squared}$. A photograph of the CALICE prototype is given in \cref{fig:CaloPicture}.

\begin{figure}
    \centering\includegraphics[width=\columnwidth]{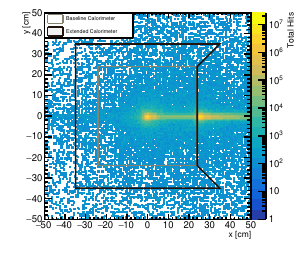}
    \caption{Estimated number of hits in an ECAL placed at a distance of \SI{3.5}{\meter} from the target for a total of $5\cdot10^{7}$ electrons on target. The photon cone in the center can clearly be seen, same as the electron beam that is diverted towards the positive x-axis. The effect of Bremsstrahlung within the recoil tracking layers is also visible by the smearing of the photon cone. The proposed approximate dimensions for the ECAL are shown for both the baseline and extended calorimeters according to \cref{sec:lohengrin:backgrounds,sec:lohengrin:sensitivity}. The number of simulated events does not correspond to the planned number of EoT. This merely serves to illustrate the size of the ECAL, the beam spot and the geometry of he setup.}
    \label{fig:CaloHitRate}
\end{figure}

The estimated rate of electrons and photons per cell that hit the calorimeter, placed at a distance of \SI{3.5}{\meter} from the target is shown in \cref{fig:CaloHitRate}, for a runtime of \SI{0.5}{\second}. The maximum hitrate per cell is roughly \SI{10}{\mega\hertz}. While this means that the average time between two subsequent hits in the same cell is about \SI{100}{\nano\second}, the most probable value is significantly shorter than that. The shaping time for the calorimeter cells must hence be sufficiently fast to allow for a high detection efficiency of high energy photons in a quasi-constant background of lower energy photons.

\begin{figure}
    \centering
    \includegraphics[width=\columnwidth]{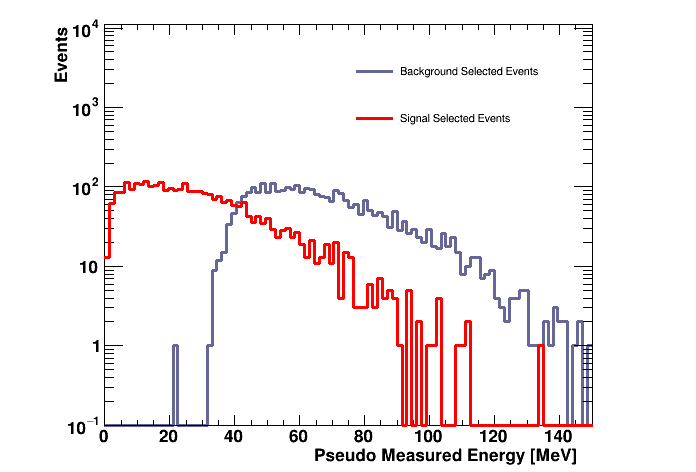}
    \caption{Calorimeter response for SM events with a high energy photon in the final state that pass the L0 trigger and for the same events where the energy deposited in the calorimeter by the photon is removed. A peaking time of $\tau=\SI{30}{\nano\second}$ has been assumed for the CRRC shaper in the analog frontend. Analog signal information was used for this plot.}
    \label{fig:ECALEnergyReadout}
\end{figure}

A feasibility study was done making simplified assumptions about the analog front-end in the calorimeter readout ASIC. For a total of $8\cdot10^{5}$ SM bremsstrahlung events, the total energy deposition in the calorimeter was simulated using \Geant, and digitised using an approximation of a symmetric CRRC shaper with a variable peaking time $\tau$ in an analog front-end with a track-and-hold readout. In this simulation pile-up, i.e.\ two or more photons hitting the same calorimeter cell before return to baseline, is taken into account. For each \lohengrin event window with a duration of $\Delta t = 2$\,ns, the number of electrons on target in this extraction cycle was randomised according to a Poisson distribution with an expectation value of $\nu = 0.2$. Events were then subject to a simulated online selection, using the proposed L0 trigger (see \cref{sec:lohengrin:detector:tdaq}). For events that occur at time $t = t_0$ passing the trigger, the readout of the calorimeter was simulated, i.e. the response of all cells at the time $t_0 + \tau$ was integrated over the full calorimeter, providing the response of the ECAL for events with low momentum electrons and high energy photons. In a second run, the same set of events was run through the same simulation, this time however removing the energy deposits from the high energy photons, emulating the response of the ECAL for signal events with low energy electrons but no high energy photons in the final state. This was done for peaking times of $\tau = (10, 30, 60, 180)\,\text{ns}$. For a peaking time of $\tau=\SI{30}{\nano\second}$ a cut that eliminates $\approx\SI{100}{\percent}$ of the SM background events, a signal efficiency of \SI{68}{\percent} can be maintained. This seems an acceptable signal efficiency and is used in the following sensitivity estimate as the baseline. The corresponding uncalibrated response from the calorimeter for such events is shown in \cref{fig:ECALEnergyReadout}. For a peaking time of $\tau =\SI{60}{\nano\second}$, a cut that would effectively remove any SM background from the signal region would reduce the signal efficiency to \SI{4.5}{\percent}. The ASIC that is implemented in the CALICE SiW ECAL prototype has a peaking time of $\tau = 180$\,ns. The requirements for the analog signal processing in the readout ASIC of a SiW calorimeter with square cells of $\num{5.5}\times\SI{5.5}{\milli\meter\squared}$ are hence rather stringent compared to the performance of the above mentioned prototype. Improvements on the current design of the ASIC are therefore required.

\subsubsection{Hadron Calorimetry}\label{sec:lohengrin:detector:hcal}

The hadronic calorimeter plays a crucial role in vetoing possible backgrounds. It will mainly be used for neutral hadrons as they will not generate a measurable signal in other parts of the detector. We expect background mainly from two types of particles: $K_L$ and $n$. Both can be created in relevant numbers and mostly traverse the detector without first decaying into partially visible decay products. 
In order to get the maximum veto efficiency, the calorimeter needs to cover the maximum solid angle possible, which is achieved by building it around the ECAL and possibly extending the coverage towards the magnet outside of the photon cone (avoiding the electron beam, however).  

Starting from the concept in \cref{sec:lohengrin:detector:ecal:energymeasurement} we implemented a simple calorimeter in our simulation framework. It consists of a sandwich of iron absorber layers (\SI{2}{\centi\meter}) and silicon layers (\SI{0.05}{\centi\meter}) as active material. This allows for an estimation of the detection efficiency.

We make some assumptions for the calculation of the veto efficiency which will be explained in the following. The particles used in simulation are $K_L$. We expect results for neutrons to be similar when adjusting for mass differences in the momentum/energy of the particles being shot at the calorimeter. The particles are shot perpendicularly at the hadronic calorimeter allowing us to cleanly study the effectiveness of different calorimeter thicknesses. We assume a particle to be detected and thus vetoed if it deposits a minimum amount of energy, $E_{\mathrm{cut}}$, in all calorimeter layers combined. 

\begin{figure}
    \centering
    \includegraphics[width=\columnwidth]{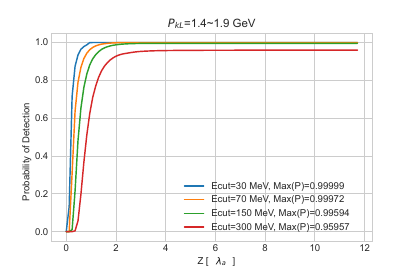}
    \caption{Expected detection probability of neutral hadrons ($K_L$) for different cutoff energies and hadron momenta considering calorimeter thickness (Z) in absorption length ($\lambda_a$).}
    \label{fig:HCalEfficiency}
\end{figure}

\cref{fig:HCalEfficiency} shows the expected veto efficiency considering the assumptions mentioned above. The sample consists of a uniform distribution of $K_L$ with \SIrange{1.4}{1.9}{\giga\electronvolt} of forward momentum. As expected the efficiency increases with increasing thickness of the calorimeter. 
We estimate the veto efficiency to be higher than 
\SI{99.999}{\percent} for a wide momentum range of neutral hadrons for a low threshold calorimeter with a thickness of more than 3 nuclear interaction lengths. In the baseline layout, we assume the HCAL to have the following dimensions: $\num{2.5}\times\num{2.7}\times\SI{1.45}{\meter\cubed}$. These dimensions might have to be revised in the future. In our simulation particles hit the HCAL perpendicularly to mitigate geometrical effects. Depending on the actual distribution of hadrons, the effective thickness of the calorimeter can be reduced if they traverse the HCAL at an angle. The hadron distribution is difficult to simulate and measurements of the hadronic background are planned (see also \cref{sec:roadmap}). The final design of the HCAL will take this into account and will have a sufficient material budget in the necessary regions of the hadron distribution. 

\subsection{Event Reconstruction}\label{sec:lohengrin:reco}
Despite the fact that many details about the detectors for the \lohengrin experiment are yet to be determined, two key ingredients to the full event reconstruction have already been studied based on a full simulation and by making some generalized assumptions about the detectors: the reconstruction and fitting of low and high momentum electron tracks in the \lohengrin tracking detector and the reconstruction of electromagnetic clusters from photons in the ECAL.
\subsubsection{Track Fitting}\label{sec:lohengrin:reco:tracking}
Track fitting is a key component of the proposed experiment since the triggering and background rejection strategy rely heavily on tracking results.

Due to the smallness of the electron mass, track reconstruction for electrons is more complicated than for most other particles. As many other experiments, we employ a Gaussian Sum Filter (GSF), i.e. a combination of several weighted Kalman Filters, for the reconstruction of the electron tracks in \lohengrin. 

\begin{figure}[ht]
    \includegraphics[width=\columnwidth]{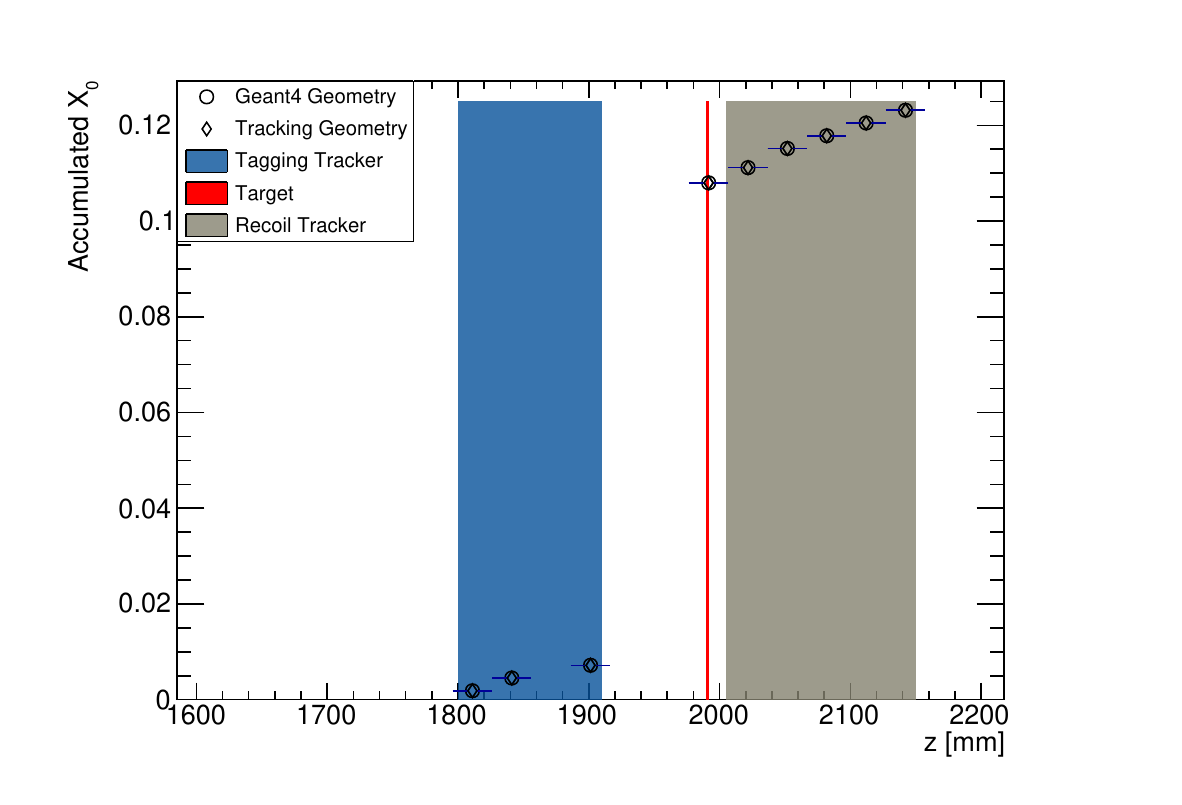}
    \caption{Material along the propagation direction as it is used in \Acts. Compared is the material encountered by just reading the GDML with full information about the detector construction and the material as it is described in the tracking geometry and thus also used for the resulting track fit.}
    \label{fig:MaterialInActs}
\end{figure}

We have implemented our tracking through A common tracking software (\Acts) \cite{ACTSProjectPaper}, a toolkit designed to provide high level reconstruction algorithms which are agnostic to detector layout and magnetic field configurations. \Acts provides a GSF that we have applied to evaluate the key performance parameters of a candidate \lohengrin tracker wrapped around a \Geant simulation. Each detector plane consists of a sandwich of \SI{100}{\micro\meter} of silicon and \SI{1000}{\micro\meter} of polyethylene. This allows us to model the effects of ultra thin sensors with silicon as the active area. The polyethylene is used to account for a wide range of support structures needed for operation of the chips. As described in \cref{sec:lohengrin:detector:trackerDetailed}, we assume square pixels with a pitch of \SI{33.04}{\micro\meter}. The planes are then positioned at the coordinates given in \cref{tab:TrackerPositions}.
\cref{fig:MaterialInActs} shows a comparison of the material encountered by the electrons along the propagation direction. ``Geant4 geometry" denotes the material accumulated by reading a GDML file description with full detector information and then subsequently running a \Geant simulation and recording the traversed material. ``Tracking Geometry" stands for material encountered along the propagation direction through the constructed tracking geometry\footnote{\Acts uses a simplified version of the entire detector for performance reasons. This version is referred to as tracking geometry. \cref{fig:MaterialInActs} shows both distributions as a validation.}. The total material budget matches well between the standalone \Geant and the \Acts implementation, as is expected due to the relatively simple description of the pixel layers in the standalone \Geant implementation.
We thus have the ability to correctly account for material effects within the tracking geometry.

\begin{table*}[t]
\caption{Positions of the middle of the silicon part of the tracking layers used as well as the middle of the target for the \lohengrin detector geometry.}
    \label{tab:TrackerPositions}
    \centering
    \begin{tabular}{cccccccccccc}
    \toprule
       Layer  & Target & 1 & 2 & 3 & 4 & 5 & 6 & 7 &8 & 9\\
       \midrule
       z Position $[\si{\milli\meter}]$ & \num{2000} & \num{1810} & \num{1840} & \num{1900} & \num{2010}& \num{2030}& \num{2045}& \num{2070}& \num{2100} & \num{2130}\\
       \bottomrule
    \end{tabular}
\end{table*}

The main challenge of this experiment considering the trigger strategy is tracking and reconstruction of low momentum electrons behind the target with acceptable performance. We thus mainly focus on this momentum range going forward.

We use an electron particle gun within the \Geant simulation in \Acts and fit the resulting hits with the implementation of the GSF. Seeding is done via the \texttt{TruthEstimated} algorithm. This takes truth information into account for pattern matching. Since we expect mostly single electrons per event, we consider this to not introduce unrealistically high seeding efficiencies. The track fit parameters are subsequently determined without taking truth information into account.

\begin{figure}
    \includegraphics[width=\columnwidth]{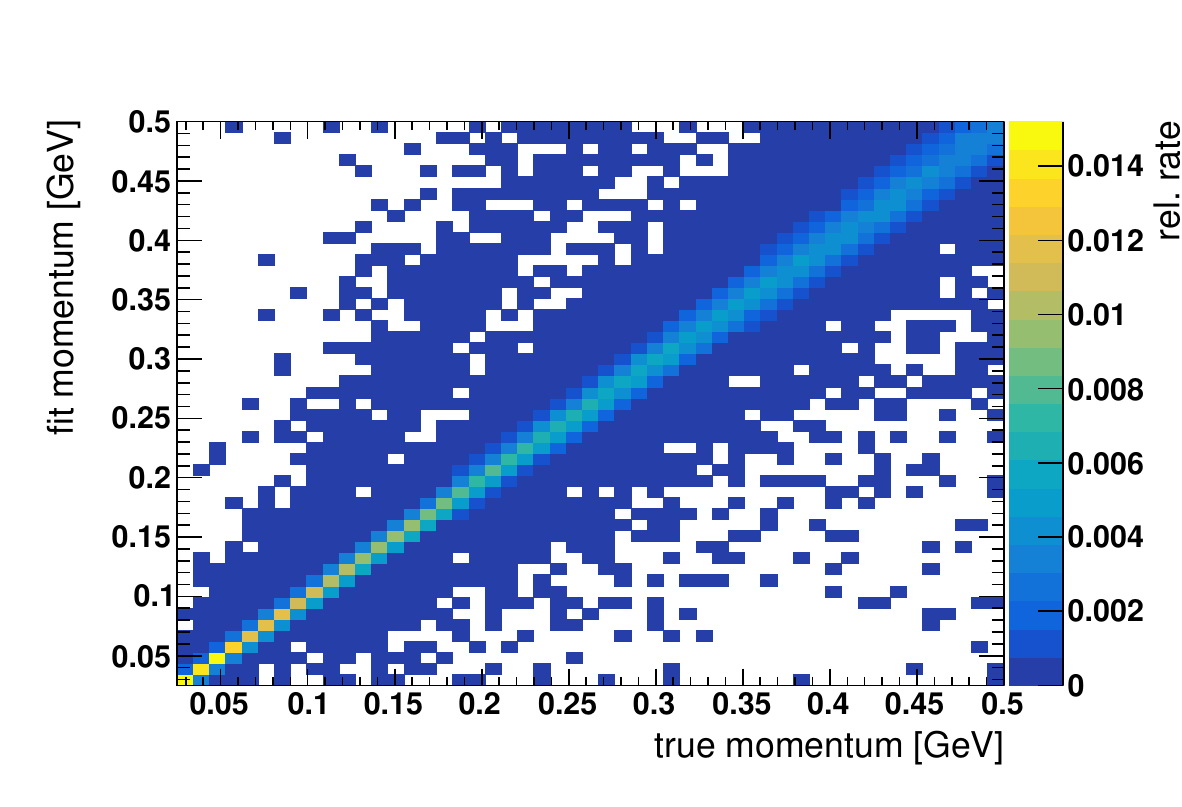}
    \caption{Comparison of fitted and true momentum.}
    \label{fig:fitVsTrueMomentum}
\end{figure}

\begin{figure}
    \includegraphics[width=\columnwidth]{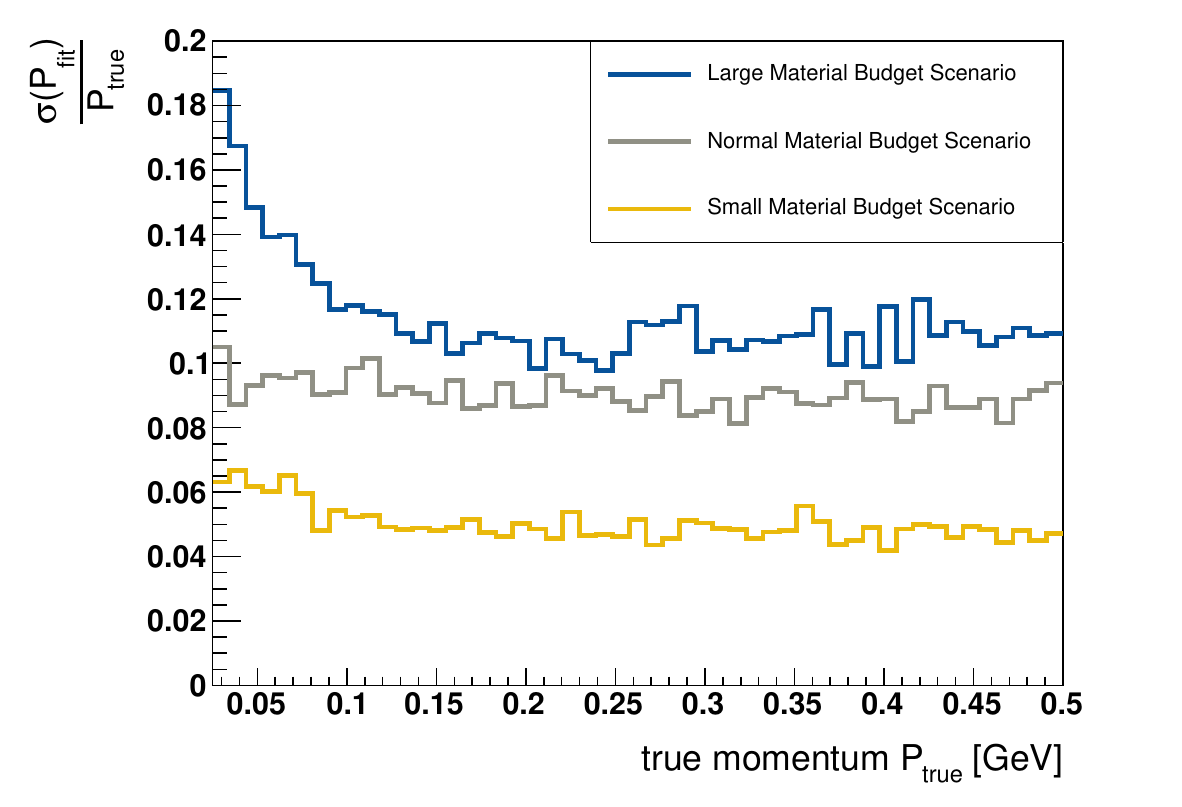}
    \caption{Relative resolution of fitted momentum from true momentum for different true momenta and possible tracker configurations considering the total material budget. The average deviation was determined using the average of a sample for each true momentum bin. The legend is to be interpreted as follows: \SI{400}{\micro\meter}(\SI{100}{\micro\meter}) silicon and \SI{1000}{\micro\meter} polyethylene for the large (normal) material budget scenario and \SI{50}{\micro\meter} Silicon and \SI{0}{\micro\meter} polyethylene for the small material budget scenario.}
    \label{fig:ProjectionDeviationVsTrueMom}
\end{figure}

In order to estimate the performance of the recoil tracker only, an electron sample with a uniform energy distribution from \SIrange{25}{500}{\mega\electronvolt} passing through a constant field of \SI{0.9}{\tesla} is generated and analysed. 
 We show a variety of performance plots for the resulting fit in \cref{fig:fitVsTrueMomentum,fig:ProjectionDeviationVsTrueMom}, making use of the six tracking layers of the recoil tracker only.

It can be seen that the fit works well in general. The fitted momentum correlates nicely with the true momentum as shown in \cref{fig:fitVsTrueMomentum}. The accuracy of the fit will be used in \cref{sec:lohengrin:physics} for an overall sensitivity estimate.

Note that there is a potential to fake signal events from reconstructed electrons with a large difference to the actual true momentum (regions far off the diagonal in \cref{fig:fitVsTrueMomentum} ). The performance shown in \cref{fig:fitVsTrueMomentum} contains single electrons fitted with a GSF without any quality cuts applied. A quality criterion is difficult to define for a GSF. However using a combination of eliminating tracks with a large residual in at least one of the tracking layers and large differences in reconstructed momentum in the first and last tracking layers, it is possible to remove these outliers to a negligible level while preserving more than \SI{97}{\percent} of tracks. The outliers can ultimately be traced back to significant bremsstrahlung events in the first layers of the tracker. Hence impairing the seeding algorithm in finding a suitable set of starting parameters for the fit. It is understood that a more sophisticated and specialised seeding algorithm might alleviate this problem without the need for further quality cuts. 


It is also worth mentioning the resolution
of the fitted momentum depicted in \cref{fig:ProjectionDeviationVsTrueMom}. This gives an idea about the general capabilities of our reconstruction chain as it is implemented here. We also show this figure of merit for two other scenarios representing different material budgets for the tracking planes. In the large material budget scenario, the thickness of the silicon layer is increased to \SI{400}{\micro\meter} and in the small material budget scenario the thickness of the silicon layer is halved and the polyethylene removed (\SI{50}{\micro\meter} and \SI{0}{\micro\meter
} respectively) compared to the normal scenario. This allows for a glimpse into an extreme case where it is possible to remove most support structures and obtain useful results with a very thin monolithic detector. It is especially useful to compare it to a rough estimate of the performance which should be achieved theoretically. Such an estimate can be found in \cite{GlucksternMeasurementErrors}.

In the momentum range we consider here, the resolution is dominated by limitations imposed by multiple scattering. We can hence neglect the track measurement term for the total momentum resolution. 
For trajectories traversing five perfectly efficient tracking layers 
and a length of \SIrange{7}{10}{\centi\meter} for the tracking assembly, the expected momentum resolution is $\approx \SIrange{8}{10}{\percent}$ for the normal scenario. The other scenarios should yield about \SIrange{11}{17}{\percent} (\SIrange{3}{5}{\percent}) in case of large (small) material budget trackers. 

\cref{fig:ProjectionDeviationVsTrueMom} shows the achieved momentum resolution as a function of the track momentum in full simulation, which roughly matches the expectation. It should be noted that \cref{fig:ProjectionDeviationVsTrueMom} includes the tracks of a large number of electrons that hit less than 5 tracking layers, in particular in the low momentum regime. The impact of multiple scattering on the momentum reconstruction is clearly visible for low momentum tracks in particular for the larger material budget scenario\footnote{We point out, however, that the used track reconstruction algorithm was not optimised for low momentum electron tracks. It may be possible to enhance the momentum measurement for low energy electrons by tuning the used GSF for such electrons.}. 

\begin{figure}
    \centering
    \includegraphics[width=\columnwidth]{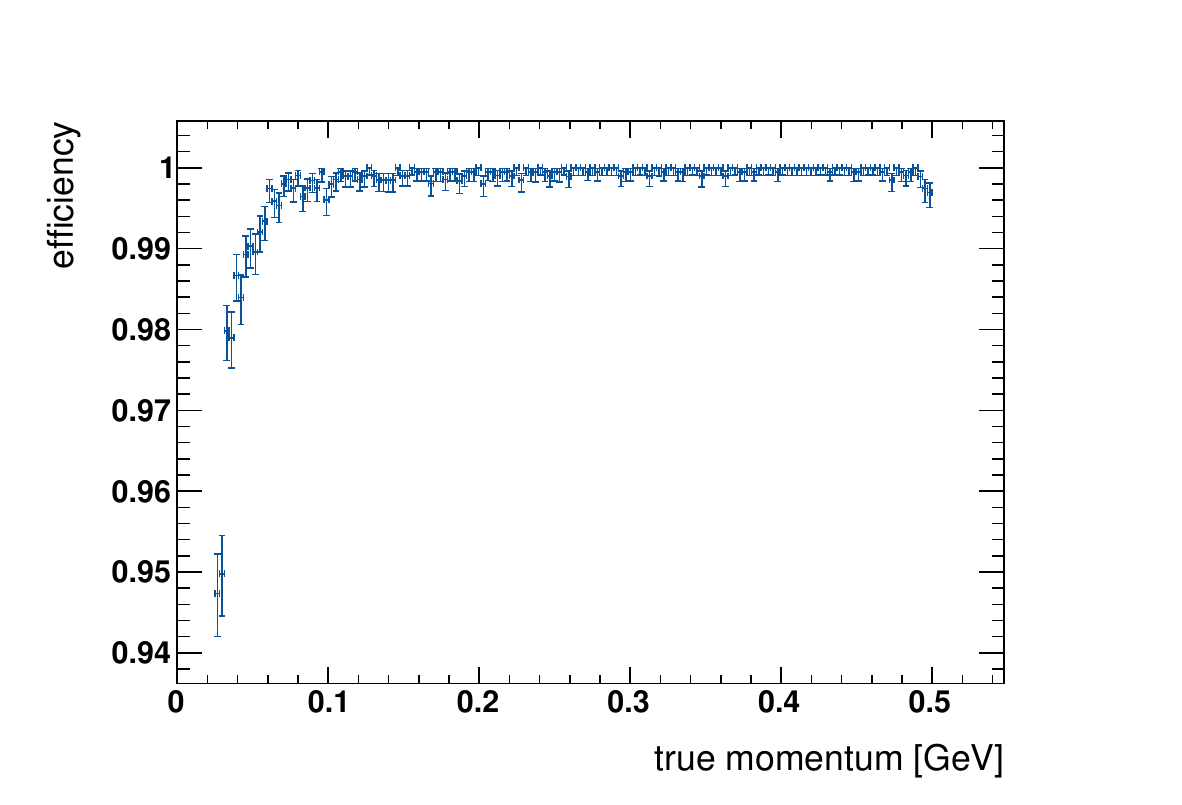}
    \caption{Tracking efficiency for different momenta in the low momentum regime. We obtain a near constant efficiency with the described setup.}
    \label{fig:TrackingEfficiency}
\end{figure}

Another key parameter for the \lohengrin experiment is the track reconstruction efficiency. 
The tracking efficiency that is obtained by the above described setup is shown in \cref{fig:TrackingEfficiency}.
It is reasonably high across the range of low momenta, falling off steeply however towards the lower boundary of the interval of interest (\SI{25}{\mega\electronvolt}). This emphasizes one of the main challenges for the proposed experiment - the efficient reconstruction of electrons down to the lowest possible energy. This is also assumed to improve as further work is done to optimize the tracker positioning and size to the specific needs of the experiment.

\subsubsection{Calorimeter Clustering and Energy Measurement}
\label{sec:lohengrin:detector:ecal:energymeasurement}
The second cornerstone of the \lohengrin event reconstruction chain is the identification of high energy photons in the electromagnetic calorimeter. In order to estimate the expected performance, in addition to the requirements on the analog signal processing that are discussed above, a simple clustering algorithm is used to build clusters from hit segments. The deposited energy is then estimated by counting the number of hit segments in a cluster.

\begin{figure}
    \centering
    \includegraphics[width=\columnwidth]{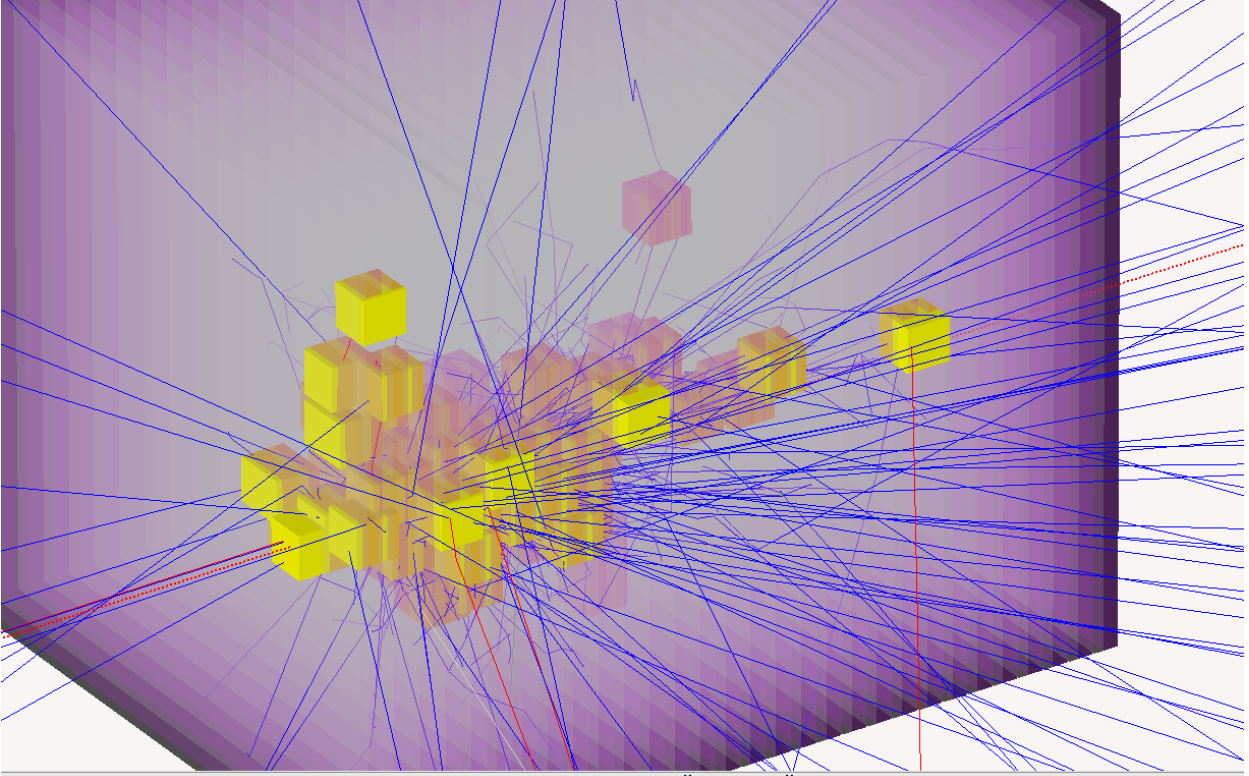}
    \includegraphics[width=\columnwidth]{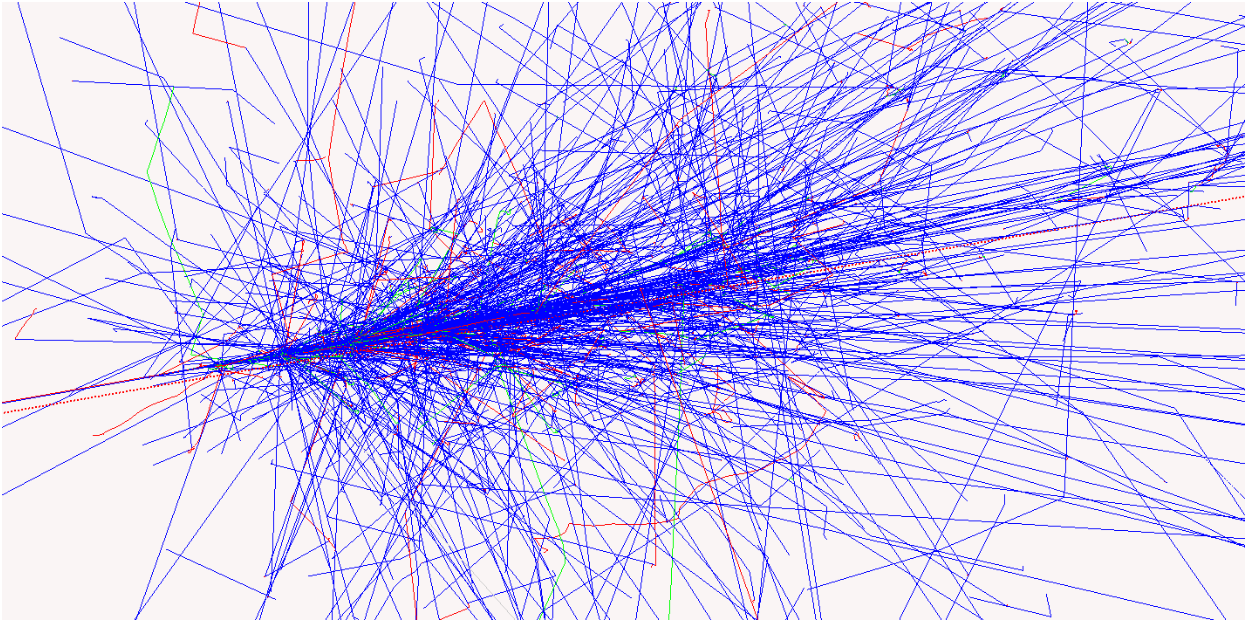}
    \caption{Formation of particle showers in a simulation of the electromagnetic calorimeter. Top: Particle shower with rendered geometry. Cubes represent digitized hits inside the sampling layers of the calorimeter. These are used for an energy estimate. Bottom: A particle shower without rendered geometry. This allows for a clearer observation of shower development.}
    \label{fig:EcalShower}
\end{figure}

We implemented the calorimeter described in \cref{sec:lohengrin:detector:ecal} in our simulation framework using \Geant for the simulation of particle interactions. The formation of particle showers inside the electromagnetic calorimeter can be observed in \cref{fig:EcalShower}.
Clustering is done in such a way that adjacent hits have to be found. Any hit counts as adjacent to another hit if it is less than a configurable number of pads away. Resulting in any pads surrounding another pad and pads in the next sampling layer at the same position counting as adjacent, if the distance is configured to be one pad. 

Each hit can then be viewed as seed for a cluster. Any adjacent hits will then be added to the cluster until no digitized hits remain and all clusters have been formed.

We have chosen a distance of one for the following analysis of energy measurements. We do however point out that more sophisticated algorithms and copious tuning potential exists. The results shown here only serve the purpose of demonstrating the potential of the selected technology in terms of energy resolution. 

In order to obtain an estimate of energy reconstruction with this setup and also the feasibility of our simulation capabilities, we use the algorithm described above to generate clusters in the electromagnetic calorimeter for different beam energies. The energy in the electromagnetic calorimeter is then reconstructed by counting the cluster size, i.e.\ the number of digitized hits. We then use the knowledge of beam energies from the simulation to perform a simple linear calibration. This transforms the number of digitised hits to an energy measurement.

\begin{figure}
    \centering
    \includegraphics[width=\columnwidth]{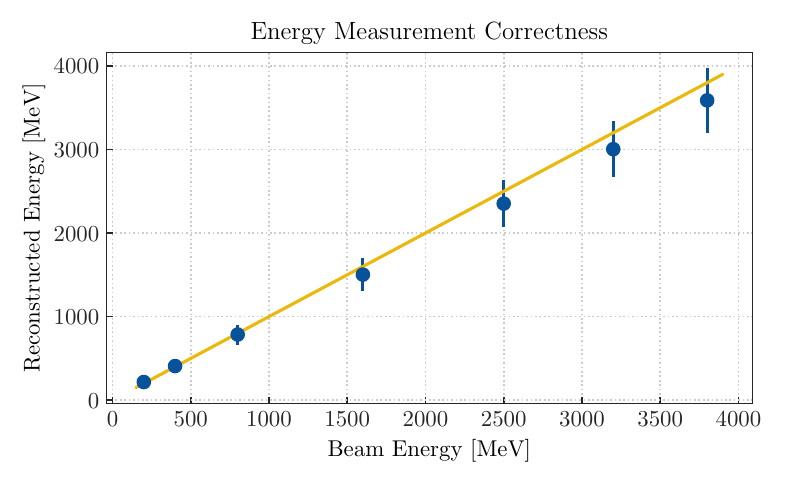}
    \caption{Reconstructed beam energy against true beam energy in the electromagnetic calorimeter after the linear calibration. The solid line represents a line through the origin with unit slope.}
    \label{fig:RecoEnergyVsTrueEnergy}
\end{figure}

\begin{figure}
    \centering
    \includegraphics[width=\columnwidth]{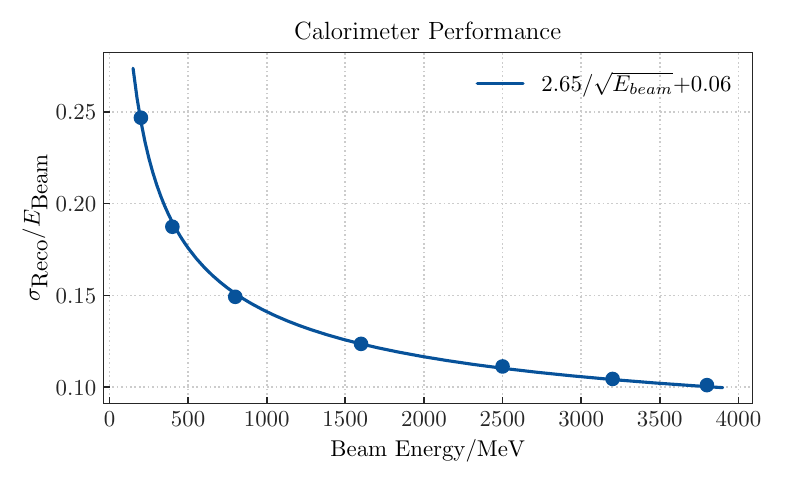}
    \caption{The ratio of energy reconstruction accuracy to beam energy with corresponding fit}
    \label{fig:CaloPerfomance}
\end{figure}

This linear calibration is then applied to every cluster and compared to the true value from the simulation. We choose a subset of specific energies and simulate several thousand events for each. Uncertainties in the energy measurement are estimated by looking at the spread of measured energies. We obtain the result displayed in \cref{fig:RecoEnergyVsTrueEnergy}. We also show the result of the ratio of energy reconstruction resolution to true beam energy in \cref{fig:CaloPerfomance}.

The linear calibration exhibits a trend towards reconstructed energies being slightly too low for larger beam energies, which could be alleviated by a more sophisticated calibration\footnote{In fact, we use additional analog signal information from the calorimeter cells for the actual event veto due to timing reasons - this is described in \cref{sec:lohengrin:sensitivity}.}. 

The ratio of the reconstructed energy and the true energy is expected to follow a function of the kind
\begin{equation*}
    \frac{\sigma_{\text{Reco}}}{E_{\text{Beam}}} = \frac{a}{\sqrt{E_{\text{Beam}}}}+b,
\end{equation*}
which is fulfilled with the algorithm described here.
The CALICE prototype has a signal to noise ratio of at least \num{10} \cite{PoeschlCalicePrototypeStatus}. We therefore neglect a possible noise term at this point as it is less relevant than the other contributions. A more detailed study will be performed in the future (see also \cref{sec:roadmap}).
The simple veto that is based on the integrated calorimeter response for events with low energy electrons in the final state can possibly be improved upon.

%% file: ExpectedPhysicsReach.tex
The discovery potential of \lohengrin is estimated using a simple cut based counting analysis. In a first step, the experiment layout is optimised, taking into account geometrical considerations in order to estimate the potential contamination of the signal region with SM bremsstrahlung events where the hard photon is missed. Provided that with the overall approach for the \lohengrin experiment this is expected to be the dominant background, other sources for backgrounds, in particular rare events with neutral hadrons in the final state, are neglected in this step. In a second step, these additional backgrounds are studied in more detail in order to estimate the overall sensitivity of the proposed experiment.

\subsection{Layout Optimisation}\label{sec:PhysicsReach:LayoutOptimisation}
\begin{figure}[ht!]
\centering
\includegraphics[width=0.45\textwidth]{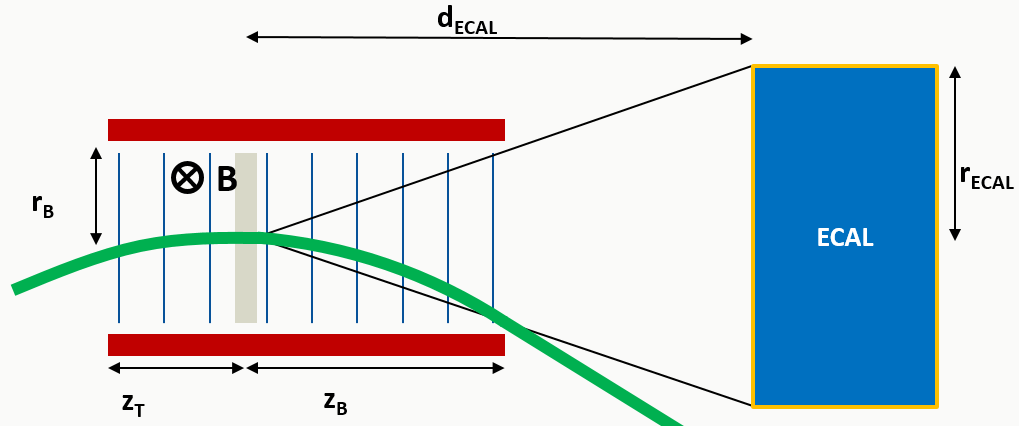}
\caption{Illustration of key parameters of the \lohengrin experiment.}
\label{fig:OptimisationParameters}
\end{figure}

\begin{figure}[ht!]
\centering
\includegraphics[width=0.49\textwidth]{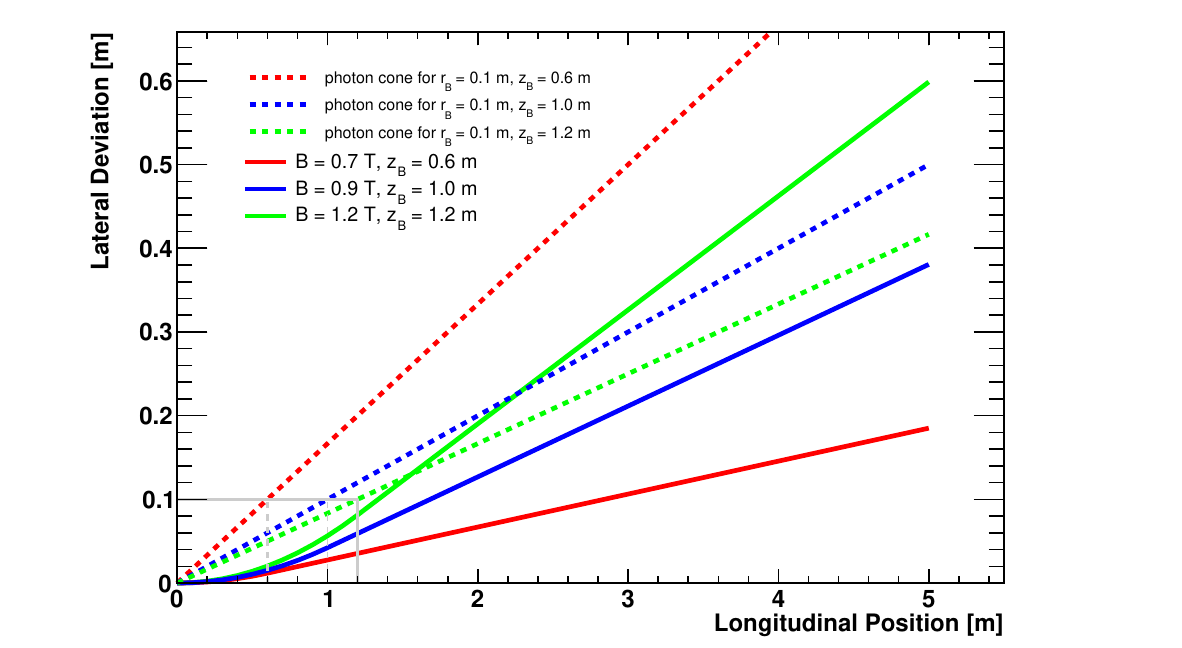}
\caption{Trajectories for 3.2\,GeV electrons that emerge from the target at a polar angle of $\theta = 0$ and $x = y = 0$. The x-axis shows the longitudinal position behind the target (z), the y-axis shows the lateral deviation due to the deflection of the electron in the magnetic field. The dashed lines indicate the cones that an ECAL would ideally cover to measure the energy of any photon that is emitted in the target and whose trajectory does not intersect with the magnet. The physical dimensions of the magnets are indicated as thin gray lines in the bottom left of the plot.}
\label{fig:LohengrinSetup:MagnetImpact}
\end{figure}

The principle strategy for \lohengrin relies on an efficient track-based reconstruction of the scattered electrons behind the target without the use of any energy measurements in the ECAL. Instead the layout features a magnetic field behind the target, strong and long enough to bend the electron beam around the ECAL. The ECAL must be large enough to allow for the reconstruction of photons that are emitted from the target in an angle as large as possible. These requirements are partially conflicting: 

The lateral dimensions of the magnet are constrained by the necessity to have a rather homogeneous and strong magnetic field. The required bending power can only be achieved through a sufficient length of the magnet, which limits the maximum polar angle for photons that are radiated off the electrons in the target and reach the ECAL. This maximum polar angle determines the maximum reasonable lateral size of the ECAL at a given distance behind the target. The key parameters of the proposed setup are shown in \cref{fig:OptimisationParameters}. The six parameters are:
\begin{itemize}
    \item $r_B$: the lateral size of the aperture of the magnet.
    \item $z_T$: the longitudinal position of the target inside the magnet.
    \item $z_{B}$: the length of the magnet behind the target.
    \item $B$: the maximum strength of the magnetic field.
    \item $d_{\textrm{ECAL}}$: the distance between the target and the ECAL.
    \item $r_{\textrm{ECAL}}$: the lateral size of the ECAL, which is assumed to be square with a side length of \mbox{$2\cdot r_{\textrm{ECAL}}$}. Based on the existing designs for the CALICE SiW prototype, $r_{\textrm{ECAL}}$ is only varied in steps of \SI{8}{\centi\meter}, as one ASU~\cite{Kawagoe_2020} has a size of $\sim \num{16}\times\SI{16}{\centi\meter\squared}$. A more complex shape of the calorimeter is discussed in \cref{sec:lohengrin:backgrounds,sec:lohengrin:sensitivity}.
\end{itemize}
The layout of the \lohengrin experiment has been optimised following a simple approach in three different scenarios: \\
In a first step, basic assumptions have been made about the magnet. A magnet similar in design to those used in the FASER experiment \cite{FASER:2018bac}
is considered, with variable assumptions about the strength of its magnetic field. Three different configurations are taken into account:
\begin{itemize}
\item \textbf{pessimistic:} $B = \SI{0.7}{\tesla}$, $z_{B} = \SI{0.6}{\meter}$
\item \textbf{baseline:} $B = \SI{0.9}{\tesla}$, $z_{B} = \SI{1.0}{\meter}$
\item \textbf{optimistic:} $B = \SI{1.2}{\tesla}$, $z_{B} = \SI{1.2}{\meter}$
\end{itemize}
For all three  configurations, $r_B = \SI{0.1}{\meter}$ and $z_{T} = \SI{0.4}{\meter}$ is chosen. A simplistic comparison of the three magnet scenarios is provided in \cref{fig:LohengrinSetup:MagnetImpact}. This figure shows the trajectories of 3.2\,GeV electrons that emerge from the target at $x = y = z = 0$ at a polar angle of $\theta = 0$ in the respective magnetic field, which is assumed to be 100\% of the nominal strength inside of the magnet and 0 outside of the magnet. In addition, the figure indicates the cones for photons that emerge from the target and do not collide with the magnet. Ideally, an ECAL would cover the full photon cone with all electrons missing the ECAL. \cref{fig:LohengrinSetup:MagnetImpact} shows that this is only achievable in the optimistic magnet scenario - for the baseline scenario and the pessimistic scenario, the bending power of the magnetic field is not enough to cover the full photon cone without a significant fraction of the beam electrons hitting the ECAL as well. A configuration as such in the pessimistic or baseline scneario may still be acceptable, provided that the number of QED bremsstrahlung events that are misidentified as signal candidates due to the missing photon veto is not too high.

\begin{figure}[ht!]
\centering
\includegraphics[width=0.45\textwidth]{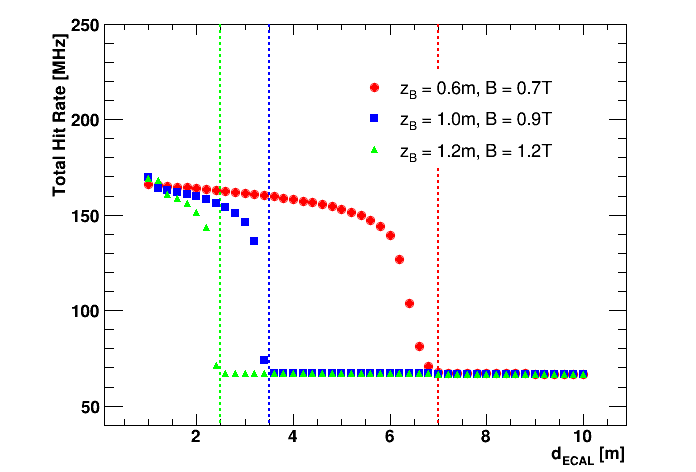}
\caption{Expected ECAL hit rates as a function of the distance between the target and the ECAL for a lateral size of the ECAL of $r_{\textrm{ECAL}} = 0.24\,$m for the three basic magnet scenarios described in the text.}
\label{fig:OptimisationECALRates}
\end{figure}

For a more quantitative analysis, and the determination of the two remaining parameters, $d_{\textrm{ECAL}}$ and $r_{\textrm{ECAL}}$, the size of the beamspot on target is taken into account: the electron beam is assumed to hit the target at a 90$^\circ$ angle with a Gaussian beam profile in both lateral dimensions, centered around $x = y = 0$ with a standard deviation of $\sigma_{x,y} = \SI{1}{\milli\meter}$. 
Any photons that are emitted at an angle $\theta_{\gamma, \textrm{max}} < \arctan\frac{r_{\textrm{B}}}{z_{\textrm{B}}}$, and with an energy of $E_{\gamma} > \SI{10}{\mega\electronvolt}$ are assumed to be measured in the ECAL as long as $\theta_{\gamma} < \arctan \frac{r_{\textrm{ECAL}}}{d_{\textrm{ECAL}}}$. \\
In order to keep the size of the ECAL as small as possible, for each of the three scenarios described above it is placed as close to the target as possible, while keeping the total rate of electrons and photons with an energy of more than \SI{10}{\mega\electronvolt} at a minimum. The expected total hit rate as a function of the distance for a lateral size of the ECAL of $r_{\textrm{ECAL}} = \SI{0.24}{\meter}$ is shown in \cref{fig:OptimisationECALRates}. The expected rate is calculated for SM events counting only electrons and photons with an energy of more than 10\,MeV whose trajectories intersect with the ECAL; the hit rates from other particles are significantly lower than the rates for electrons and photons and can be neglected for the purpose of this study. For any given lateral rate of the ECAL, the hit rate decreases for increasing $d_{\textrm{ECAL}}$ as less electrons and photons hit the ECAL. The expected hit rate approaches a quasi-constant pedestal of more than 60\,MHz for large distances - this residual hit rate is almost exclusively due to photons that are emitted at small polar angles in the target and are not affected by the magnetic field. For the baseline scenario with a moderately dimensioned magnet system, placing the ECAL at a distance of \SI{3.5}{\meter} from the target meets the goal of avoiding a flooding of the ECAL with the primary electron beam while achieving a reasonable coverage for photons. For comparison, a benchmark with a smaller and a benchmark with a larger ECAL have been considered in addition to the baseline scenario. A summary of the key parameters for all three scenarios and all three ECAL benchmarks is given in \cref{tab:ECALComparison}. The three options for the lateral size of the ECAL are based on the size of existing planes for the CALICE SiW ECAL prototype, as discussed above.

\begin{table*}[ht!]
\caption{Distance between the target and the ECAL for the three magnet scenarios and different choices for the lateral size of the ECAL. The distance $d_\textrm{ECAL}$ is chosen such that the total hit rate for the ECAL is reduced to its minimum for each scenario.}
\label{tab:ECALComparison}
\centering
\begin{tabular}{lccc}
\toprule
Scenario & $r_{\textrm{ECAL}} = \SI{16}{\centi\meter}$ & $r_{\textrm{ECAL}} = \SI{24}{\centi\meter}$ & $r_{\textrm{ECAL}} = \SI{32}{\centi\meter}$ \\
\midrule
pessimistic & \SI{5.0}{\meter} & \SI{7.0}{\meter} & $>$ \SI{10}{\meter} \\
baseline & \SI{2.7}{\meter} & \SI{3.5}{\meter} & \SI{5.3}{\meter} \\
optimistic & \SI{2.2}{\meter} & \SI{2.5}{\meter} & \SI{3.5}{\meter} \\
\bottomrule
\end{tabular}
\end{table*}
The distances between target and ECAL that are shown in the table are the minimum distances that are required in order to keep the total hit rate in the ECAL at a manageable level. The higher the bending power of the magnet, the closer to the target the ECAL can be placed, reducing the lateral size while maintaining the sensitivity for photons radiated in the forward direction. For the baseline scenario, an ECAL with a lateral size of $\num{48}\times\SI{48}{\centi\meter\squared}$ would have to be placed at a distance of \SI{3.5}{\meter} from the target in order to minimise the number of electrons that hit the ECAL. Considering the available experimental area at the ELSA accelerator and the fact that the ECAL must be embedded in a significantly larger HCAL, this is viable. A symmetric, larger ECAL would enhance the coverage for photons escaping the magnet in the forward direction to almost \SI{100}{\percent}; however, it would also catch a significant fraction of the primary beam if placed at the same distance from the target. Placing the ECAL at a larger distance from the target is not feasible given the spatial constraints at ELSA. In addition, the coverage in terms of the photon polar angle would be significantly diminished. A smaller ECAL could be placed much closer to the target in the baseline scenario. However, it would significantly limit the angular coverage of the ECAL, reducing the sensitivity of the experiment. Hence the baseline for the ECAL position and lateral size is set to $d_{\textrm{ECAL}} = \SI{3.5}{\meter}$ and $r_{\textrm{ECAL}} = \SI{0.24}{\meter}$.

\begin{figure}[ht!]
\centering
\includegraphics[width=0.45\textwidth]{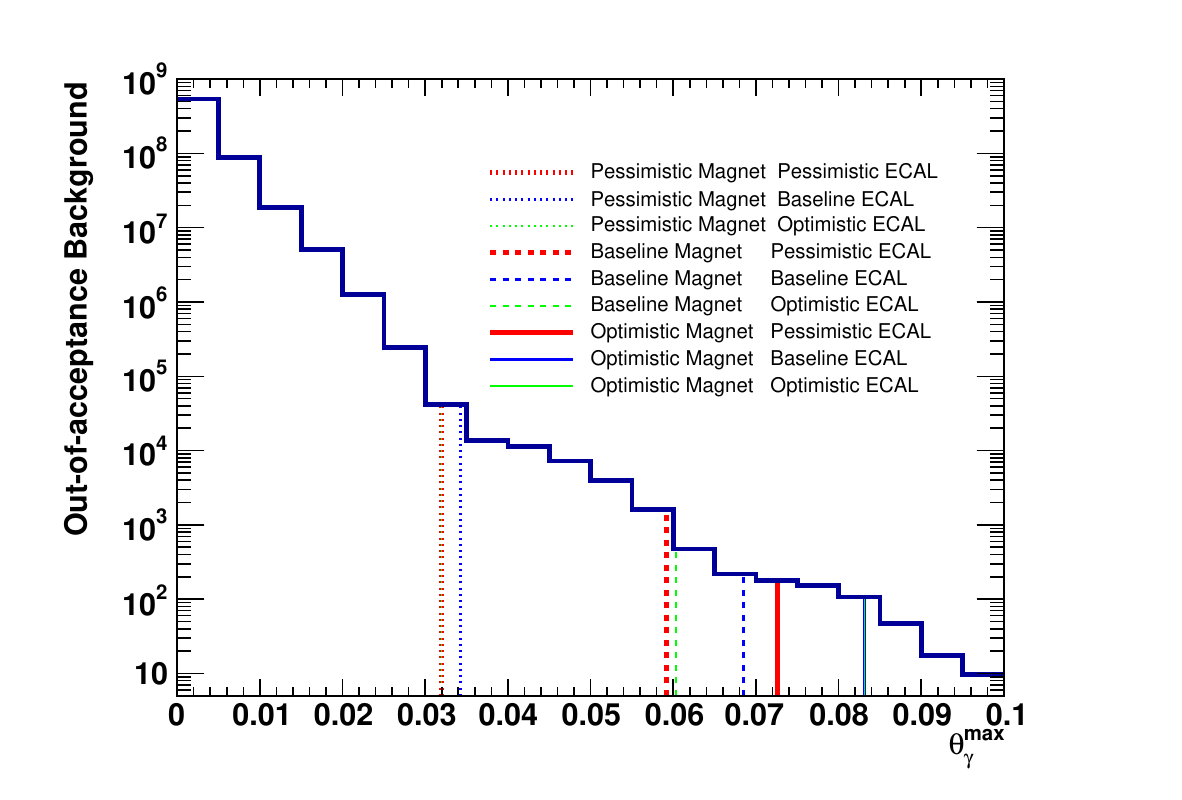}
\caption{Expected number of SM events with a low energy electron in the final state and a photon that is emitted at angles larger than $\theta_{\gamma}^{\textrm{max}}$. The dotted/dashed/solid lines show this maximum angle for the pessimistic/baseline/optimistic magnet scenarios in red/blue/green for the pessimistic/baseline/optimistic ECAL scenarios.} 
\label{fig:Lohengrin:PolarPhotonAcceptance}
\end{figure}

The expected number of events with a low energy electron ($p_e < 75$\,MeV) and a photon that is emitted from the target at a polar angle that is larger than some maximum acceptance $\theta_{\gamma}^{\textrm{max}}$ is shown in \cref{fig:Lohengrin:PolarPhotonAcceptance}. The figure also shows the maximum acceptance angle for the different magnet and ECAL scenarios that have been discussed above. A few scenarios have identical limits, e.g. the baseline and optimistic ECAL scenarios for the optimistic magnet configuration. This is due to the fact that in these scenarios the acceptance is limited by the magnet bore rather than the actual size of the ECAL. It is clear that the photon acceptance can be significantly improved with a stronger, larger magnet, as that allows to increase the size of the ECAL without flooding the ECAL with electrons. For the remainder of this study we only consider the baseline scenario, however. This is due to the fact that preliminary discussions indicate that the baseline magnet scenario is feasible, while the feasibility of the optimistic scenario is less clear. For the baseline magnet scenario, the baseline ECAL scenario is expected to yield the highest sensitivity due to the fact that the angular coverage of the ECAL is better for this scenario than for the pessimistic and optimistic scenarios. 

As discussed above in this section, the ECAL will still be flooded with photons at a high rate. The energy of these photons is distributed according to the SM bremsstrahlung cross-section - the overwhelming majority of the photons that hit the ECAL has an energy that is considerably below the energy of the incident electrons. In events with such a low energy photon, the scattered electron is expected to retain a significant fraction of the energy of the incident electron. In order to distinguish between signal events and SM QED background, a precise energy measurement is hence not required for each event. The calorimeter must only provide a good discrimination power between events with a high energy photon that hit the calorimeter and events without such a photon. The capability of a derivative of the SiW CALICE ECAL with enhanced (faster) signal shaping in each readout channel has been discussed in \cref{sec:lohengrin:detector:ecal}. 

With a magnetic field pointing in the direction of the positive y-axis, the trajectories of electrons are bent into the direction of the positive x-axis. An extension of the ECAL in the three other quadrants in order to improve the detection efficiency for forward photons is discussed in \cref{sec:lohengrin:sensitivity}. Another option to improve the photon coverage of the ECAL and further suppress SM QED background could be the addition of a quadrupole moment to the magnet, which could be used to reduce the size of the electron beam at the location of the ECAL in one dimension, while widening it in the other lateral dimension. The feasibility and impact of such a configuration will be explored in the future. 

\subsection{Analysis Strategy}\label{sec:lohengrin:strat}
The baseline analysis for \lohengrin is a counting experiment in a predefined signal region. The signal region is defined in a way that maximises the sensitivity of the experiment for dark photons in the mass window $m_{\ap} \in [\SI{1}{\mega\electronvolt}; \SI{40}{\mega\electronvolt}]$. 
While the experiment in general and the signal region in particular are designed to maximise the rejection for SM events, rare SM processes can mimic signal events. Wherever possible, data driven methods will be used to estimate the backgrounds. Wherever that is not possible, Monte Carlo simulation is used to estimate SM backgrounds in the signal region. The implementation of the trigger system will allow for special runs that can be used to validate the background estimations in control regions that are orthogonal to the signal region.

The signal region is defined by:
\begin{itemize}
    \item The presence of at exactly one beam electron in the initial state.
    \item The presence of exactly one (baseline candidate signal region) or at most one signal electron in the final state (second candidate signal region).
    \item No significant energy deposition in the ECAL.
    \item The absence of any hadronic activity in the final state.
    \item The absence of any charged tracks other than the signal electron in the final state.
\end{itemize}
Here, a signal electron is defined as a charged track that is compatible with the electron hypothesis and meets the two selection criteria
\begin{itemize}
\item $E < E_{\textrm{e,cut}}$
\item $\theta < \theta_{\textrm{e,cut}}$
\end{itemize}
Preliminary values for the cuts are determined in \cref{sec:lohengrin:sensitivity} in order to maximise the sensitivity of the experiment in the targeted dark photon mass range. 
With the foreseen ECAL design and the high rate of electrons on target, the energy measured in the ECAL for a given, triggered event, will be subject to a substantial background pedestal from SM bremsstrahlung events as discussed in \cref{sec:lohengrin:detector:ecal}. 
The cut on the measured energy in the ECAL for events with low energy electrons in the final state is significantly higher than the average value of this pedestal in order to preserve as much signal as possible while still completely suppressing any SM background. 
Hadronic activity in the target is established through the readout of the hadronic calorimeter. The strategy for the \lohengrin experiment will not rely on a precise measurement of the hadronic energy in the event. Rather, a low noise HCAL is used to efficiently veto any events in which the total energy that is measured in the hadronic calorimeter that can not be interpreted as the extension of a shower that started in the ECAL is above a low threshold. This allows the implementation of a relatively small HCAL around the ECAL. As a last cut, the full tracking information is used again to veto any events that contain charged particles other than the signal electron emerging from the target. This way, events containing positrons from pair-production in the target, as well as muons and charged hadrons can be efficiently vetoed without relying on the ECAL and HCAL measurements. 

\subsection{Background Estimation}\label{sec:lohengrin:backgrounds}
SM backgrounds are estimated differently depending on the mechanism that is responsible for them passing the analysis cuts. Three different classes of background events are considered here - these are
\begin{itemize}
    \item out-of-acceptance backgrounds: these arise from the fact that the coverage of the detector is limited. High energy bremsstrahlung photons that are emitted at a large angle from the target can miss the ECAL and lead to a large amount of missing energy in the final state. While exceedingly rare, these comprise the dominant contribution of the expected overall background.
    \item neutral hadrons: these backgrounds arise from the production of neutral hadrons in the target, the tracking detector, or the first layers of the ECAL. Neutral hadrons can be produced in both photo-nuclear and electro-nuclear interactions. Stable neutral hadrons, mainly neutrons and long-lived kaons, can escape the experiment undetected if they either miss the HCAL completely, or deposit only very little energy while passing the HCAL. 
    \item neutrino backgrounds: in some very rare processes, neutrinos can be produced through the interaction of the final state particles with the matter in the tracking detector or the calorimeters. High energy neutrinos can lead to an energy imbalance and hence to the selection of an event in the analysis.
\end{itemize}
The layout of the experiment has been chosen such that the contribution of the out-of-acceptance background is minimised while considering reasonable choices for the ECAL dimensions, and reasonable assumptions on the strength and size of the magnet, as described in \cref{sec:PhysicsReach:LayoutOptimisation}. The out-of-acceptance backgrounds are estimated by calculating the expected number of SM bremsstrahlung events that pass the selection criteria for the signal region. Any photons with an energy above \SI{10}{\mega\electronvolt} that hit the ECAL are assumed to be registered in the ECAL, while tracks for electrons with a momentum of more than 25\,MeV are assumed to be reconstructed with high efficiency. The sensitivity is estimated by calculating the quantity $\frac{s}{\sqrt{s+b}}$ as a function of the $E_{\textrm{e,cut}}$ for $\theta_{\textrm{e,cut}} = 0.25$, which is approximately the maximum scattering angle for which electrons can still be properly tracked in the \lohengrin setup. As the out-of-acceptance backgrounds are expected to dominate the background spectrum, neutral hadronic and neutrino backgrounds have been neglected in this step, i.e. events with a large energy transfer to the target nucleus have been neglected. These events are studied in more detail below.

\begin{figure}[ht!]
\centering
\includegraphics[width=0.45\textwidth]{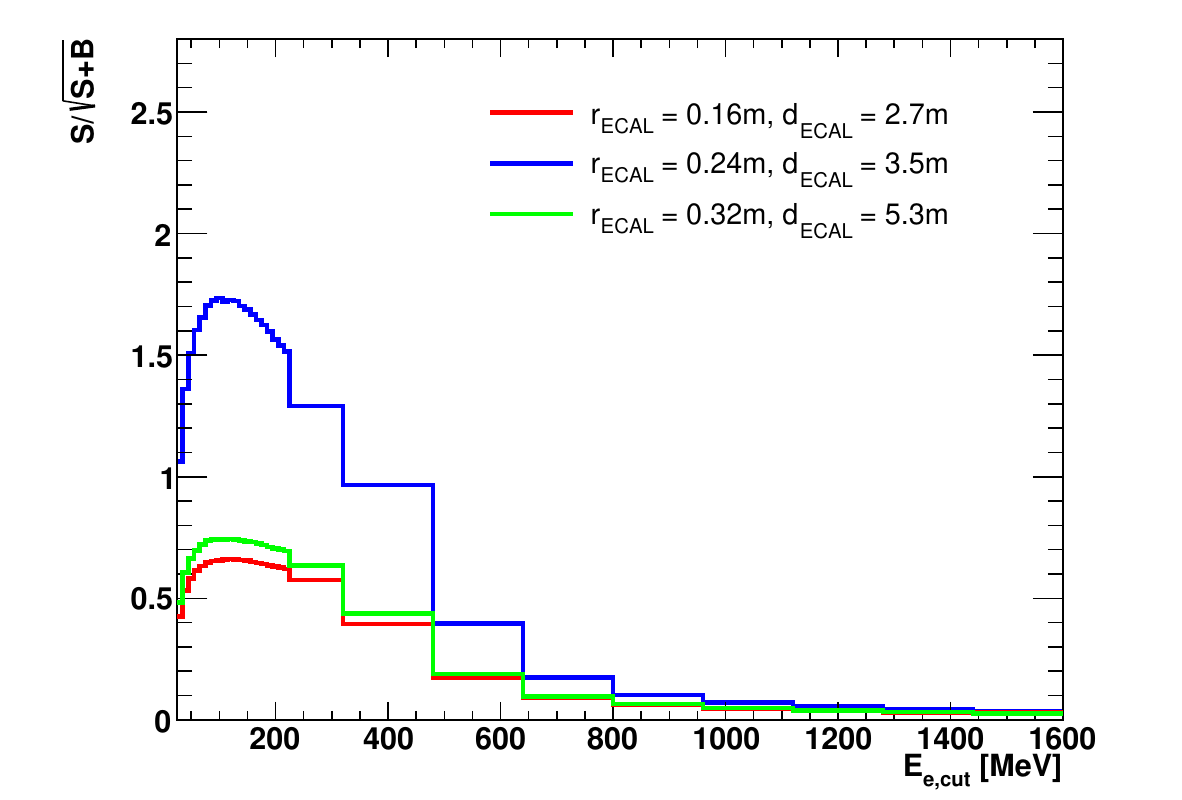}
\caption{Expected significance as a function of the upper cut on the electron energy in the baseline magnet scenario for the ECAL configurations shown in \cref{tab:ECALComparison} for a dark photon mass of $m_{\ap} = \SI{10}{\mega\electronvolt}$ and $\varepsilon = 2\cdot10^{-5}$.}
\label{fig:LohengrinCutOptimisation}
\end{figure}

The expected sensitivity for the baseline scenario as a function of $E_{\textrm{e,cut}}$ is shown in \cref{fig:LohengrinCutOptimisation} for $m_{A'} = 10$\,MeV and $\varepsilon = 2 \cdot 10^{-5}$. In addition to the upper cut on the electron energy, a lower cut of $p_e > 25$\,MeV is included in this plot in order to account for the limited reconstruction efficiency below this value. For the shown range in $E_{\textrm{e,cut}}$, any reasonable cut on the measured energy in the ECAL has no impact on the sensitivity. The figure shows a steep increase in the sensitivity with decreasing cuts on the maximum energy of the recoil electron. In the interval for $E_{e,cut} < 200$\,MeV the signal significance is improved by a factor of about $1.5$ if the signal region is extended to electrons with $m_e < E_e < 25$\,MeV, emphasizing the requirement for efficient electron reconstruction down to the lowest energy possible. An example cutflow based on the \texttt{Lohengrin++} code is shown in \cref{tab:analysis_cutflow} for SM bremsstrahlung events and three different signal benchmark points. For this study, we choose $E_{e,\textrm{cut}} = \SI{75}{\mega\electronvolt}$ and $\theta_{e,\textrm{cut}}$ = 0.25, with the additional requirement of $p_e > 25$\,MeV. As shown in \cref{fig:LohengrinCutOptimisation}, the upper cut of $E_{e,\textrm{cut}} = \SI{75}{\mega\electronvolt}$ does not maximize the sensitivity for $m_{A'} = 10$\, MeV, for which a slightly higher cut is expected to yield a higher sensitivity. For higher dark photon masses, a lower cut is however expected to yield a better sensitivity (see \cref{sec:theory}). The final cut will be determined in a more elaborate process in the future.
\begin{table*}[ht]
    \caption{Tentative Cutflow for the main background of SM photons ($\gamma$) and three benchmark points of dark photon masses and couplings. Baseline cuts include the remaining energy fraction $\xi$, electron momentum \(p_e\) and angle \(\theta_e\), and calorimeter energy \(E^{\gamma}\). The cutflow has been determined using the \texttt{Lohengrin++} code, i.e. it is based on the \textit{true} quantities of the final state particles, not measured quantities.}
    \label{tab:analysis_cutflow}
    \centering
    \begin{tabular}{lcccc}
        \toprule
         $4\cdot10^{14}$ EoT & number of $\gamma$ & $m_\ap = \SI{1}{\mega\electronvolt}$& $m_\ap = \SI{10}{\mega\electronvolt}$ & $m_\ap = \SI{100}{\mega\electronvolt}$  \\
         & & $\varepsilon = 1.2\cdot10^{-6} $ & $\varepsilon = 1.4\cdot10^{-5}$ & $\varepsilon = 1.6\cdot10^{-4}$ \\
         \midrule
         total $\xi <$ 0.95 & $3.1\cdot10^{14}$ & $26$ & $80$ & 27 \\
         $p_e < \SI{75}{\mega\electronvolt}$, $\theta_e < \SI{0.25}{\radian}$ & $1.0\cdot10^{12}$& 1.3 & 26 & 5.1 \\
         $E^{\gamma}(\theta_\gamma < 0.07) < \SI{640}{\mega\electronvolt}$ & 293 & 1.3 & 26 & 5.1 \\
        \bottomrule
    \end{tabular}
\end{table*}
For the baseline layout described in \cref{sec:lohengrin:detector}, the lateral dimension of the ECAL is of paramount importance for the level of SM backgrounds. For the baseline choice of a $\num{48}\times\SI{48}{\centi\meter\squared}$ large ECAL, a total of 293 SM bremsstrahlung background events are expected after $4\cdot10^{14}$ electrons on target. This background can be suppressed further significantly by increasing the lateral size of the ECAL in the direction of the negative x-axis as well as the positive and negative y-axes - for example, an extended ECAL that covers the full opening angle of the magnet for $|\phi| > \frac{\pi}{4}$ (see \cref{fig:CaloHitRate})
would reduce the number of SM background events in the signal region from 293 to 82 for the baseline magnet scenario.

Hadronic backgrounds are harder to estimate for several reasons. The cross sections for the relevant photo-nuclear and electro-nuclear reactions are small, and usually a large number of neutral and charged particles (mainly neutrons and protons) emerge from the reaction. Four different classes of events are considered here:
\begin{itemize}
\item electro-nuclear interaction in the target: the incident electron can transfer a large fraction of its energy onto the target nucleus, emerging from the target with a very low momentum without having radiated a real photon. In most of these interactions, the target nucleus will break up emitting one or more nucleons with a significant energy. From a simulated sample, enabling only electron-nuclear and photo-nuclear interactions and hadron decays, no event passed all signal region cuts. Due to the low recoil electron energy cut, the simulation of such events containing a signal electron is not reliable with \Geant, however. Alternatives are being investigated at this point. We assume that the number of background events with electro-nuclear interactions in the target can be well suppressed.
\item electro-nuclear interactions in the tracker or in the ECAL. With thin silicon tracking planes, the probability for an electro-nuclear interaction that leaves no significant energy in the calorimeters is very low. Such an interaction is more likely to happen in the first layers of the ECAL, which is however further suppressed by the fact that the bulk of the scattered electrons are diverted around the ECAL. In the baseline scenario, about \SI{0.5}{\percent} of all incoming electrons still hit the ECAL. As a conservative estimate, all of these electrons are assumed to have lost almost no energy in the target, and the possibility that the electron does not loose a significant fraction of its energy within the first five layers of the SiW ECAL is taken into account. As for the electro-nuclear interaction in the target, a detailed estimate is difficult to produce at this point. Out of a sample containing $10^{7}$ events, no event passed all signal region cuts. As for the electro-nuclear interactions in the target, this study gives only limited confidence that the number of background events with electro-nuclear interactions will be well controlled. However, compared to the same type of background events in the target, events with electro-nuclear interactions in the tracker or in the ECAL are suppressed by additional factors: first, if the electron-nuclear interaction occurs in the ECAL, the electron must have had a high energy and the tracking algorithm must have failed to find the track. It is rather unlikely that such an event would pass all signal region cuts. The same argument holds for interactions in the tracker. Only events where the electron-nuclear interaction occurs in one of the first two tracking planes are unlikely to be vetoed by the signal region cuts. Hence we assume that this class of events is almost negligible.
\item photo-nuclear interactions in the target and in the tracker. In a simulation of $10^{7}$ events in which electrons with an energy of $\SI{3.2}{\giga\electronvolt}$ are shot onto a \SI{10}{\percent} $X_0$ tungsten target to produce bremsstrahlung, no event with a photo-nuclear interaction passes all cuts for the signal region. 
\item photo-nuclear interactions in the first layers of the ECAL. This class of events has been simulated by studying photo-nuclear interactions of a \SI{3.2}{\giga\electronvolt} photon beam with a \SI{1}{\meter} thick tungsten target, enabling only electro-nuclear, photo-nuclear and hadronic decay processes. Any charged particles with a significant amount of energy from such interactions are assumed to leave a detectable signal in the ECAL. The detection efficiency for neutral hadrons in the HCAL is estimated to be \SI{99.999}{\percent} for neutral hadrons with an energy of more than \SI{100}{\mega\electronvolt}. Again, no event from the simulated sample of $10^{7}$ photons on target passes all signal region cuts. 
\end{itemize}

While it is difficult to estimate the number of background events that stem from photo-nuclear or electro-nuclear interactions, we assume that the overall number of background events in the signal region is largely dominated by SM bremsstrahlung events in which the photon misses the ECAL. Lacking a precise estimate for the remaining backgrounds beyond the studies outlined above, we conservatively consider a contamination of up to 10 such events in the signal region; an assumption that will be validated or improved by further studies. We consider this to be a conservative estimate compared to similar proposed experiments, i.e.\ \cite{Akesson:2018vlm}. We point out that a significantly higher number of background events would diminish the discovery potential of the proposed experiment -- if future studies indicate that it is necessary, additional veto detectors, like a reasonably sized additional HCAL that is placed outside of the magnet around the target, will be added to the \lohengrin experiment. As explained in \cref{sec:roadmap}, the experiment could be commissioned in a phased manner, allowing a measurement of photo-nuclear and electro-nuclear cross sections in the relevant recoil electron phase-space.

\subsection{Signal Efficiency}\label{sec:lohengrin:signal}
\begin{figure}[ht!]
\centering
\includegraphics[width=0.45\textwidth]{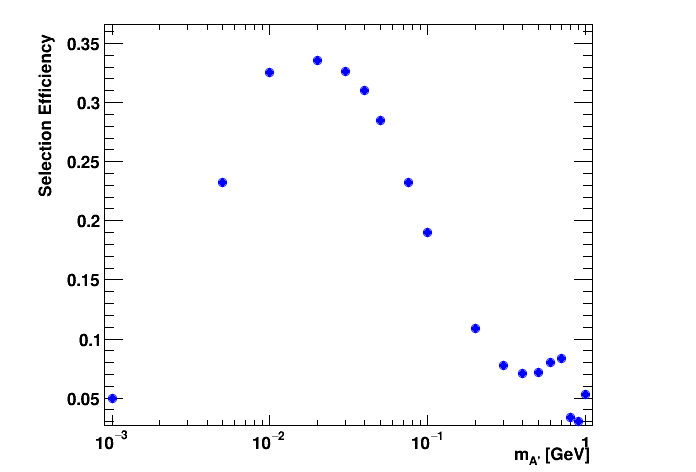}
\caption{Expected selection efficiency for a range of dark photon masses.}
\label{fig:PhysicsReach:SignalEfficiency}
\end{figure}

The selection efficiency for signal events depends on the distributions for the kinematic observables for the electron (i.e.\ the total momentum and the scattering angle). The \texttt{Lohengrin++} code has been used to calculate the expected number of dark photon events in general and in the signal region. Due to the kinematic properties of the signal events, a selection efficiency between \SI{3}{\percent} and \SI{35}{\percent} is expected for dark photon masses between $\SI{1}{\mega\electronvolt}$ and $\SI{1}{\giga\electronvolt}$. For electrons with an energy between \SI{25}{\mega\electronvolt} and \SI{75}{\mega\electronvolt}, the average trigger efficiency for $\theta_e < 0.25$ is \SI{99}{\percent}, while the average tracking efficiency is \SI{98}{\percent}. The expected selection efficiency (including triggering and tracking efficiencies) for signal events is shown in \cref{fig:PhysicsReach:SignalEfficiency}. For low dark photon masses, the efficiency is dominated by the trigger efficiency, which is low when averaged over the recoil electron energy spectrum, see \cref{fig:LohengrinSetup:TriggerEfficiencyVsE}. For high dark photon masses, the recoil electron energy spectrum is shifted to very low values, for which the reconstruction efficiency is low, see \cref{fig:TrackingEfficiency}). 

\subsection{Sensitivity Estimation}\label{sec:lohengrin:sensitivity}
The sensitivity of the \lohengrin experiment has been estimated in various scenarios. In general, two signal regions are considered: in one signal region, a reconstructed electron with an energy between \SI{25}{\mega\electronvolt} and \SI{75}{\mega\electronvolt} is required in the final state, while the energy that is measured in the ECAL must be compatible with less than \SI{640}{\mega\electronvolt} of deposited energy above the low energy photon pedestal in this event, and the hadronic veto system must not fire for this event.

In the second, more aggressive signal region, the \textit{absence} of any reconstructed electron with a momentum of more than \SI{75}{\mega\electronvolt} is required, all the other cuts hold. The second signal region significantly enhances the selection efficiency for signal events, but might lead to an unacceptable increase in the number of background events if the veto efficiency for high energy electrons is not close enough to \SI{100}{\percent}. This signal region is studied here in order to demonstrate the physics potential for the case that the lower threshold for electron reconstruction can be lowered significantly compared to the targeted threshold of \SI{25}{\mega\electronvolt}.

Two different scenarios are assumed for the ECAL as well. In the first scenario, the baseline setup is used with an ECAL that has a lateral size of $\num{48}\times\SI{48}{\centi\meter\squared}$ and that is mounted symmetrically around $x = y = 0$ at a distance of $\SI{3.5}{\meter}$. In the second scenario, the ECAL coverage is extended for 
$|\phi| > \frac{\pi}{4}$ for the $\theta < 0.1$.

\begin{figure}[ht!]
\centering
\includegraphics[width=0.45\textwidth]{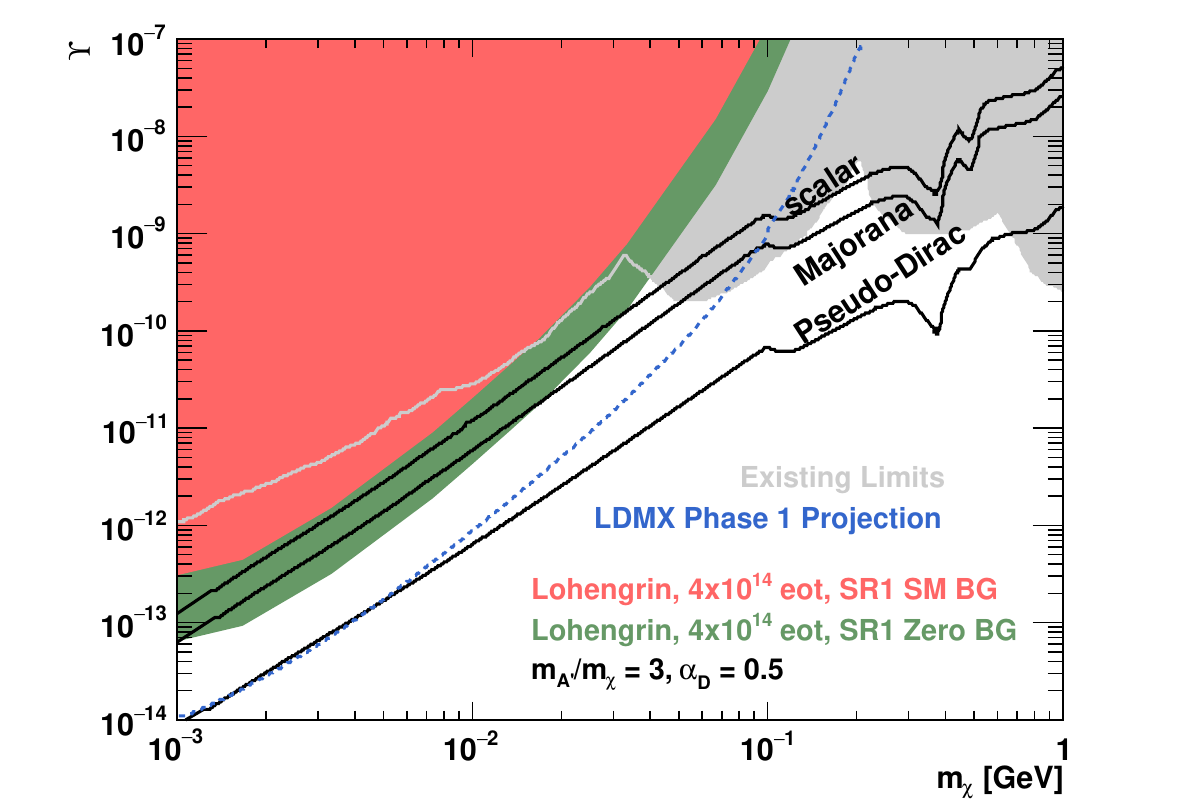}
\caption{Estimated sensitivity of the \lohengrin experiment for the extended ECAL coverage and the baseline signal electron phase space. The red area show the expected sensitivity including the estimated number of background events, the green area indicates the statistical limit for the sensitivity assuming a background free search (both limits are calculated at the 90\% CL). The gray area shows existing limits from various experiments - this information is taken from ~\cite{Graham:2021ggy} and includes limits from LEP~\cite{LEPDP2005, LEPDP2009,Fox_2011}, BaBar~\cite{Lees_2017}, MiniBooNE~\cite{PhysRevD.98.112004}, NA62~\cite{NA62_2019}, LSND~\cite{deNiverville_2011}, NA64~\cite{Banerjee_2019} as well as CRESST II~\cite{Angloher_2016}, CRESST III~\cite{Abdelhameed_2019} and XENON1T~\cite{Aprile_2019}. In blue the expected limit from Phase 1 of LDMX is shown~\cite{Akesson:2022vza}.}
\label{fig:PhysicsReach:Sensitivity_SR1_StandardECAL}
\end{figure}

\begin{figure}[ht!]
\centering
\includegraphics[width=0.45\textwidth]{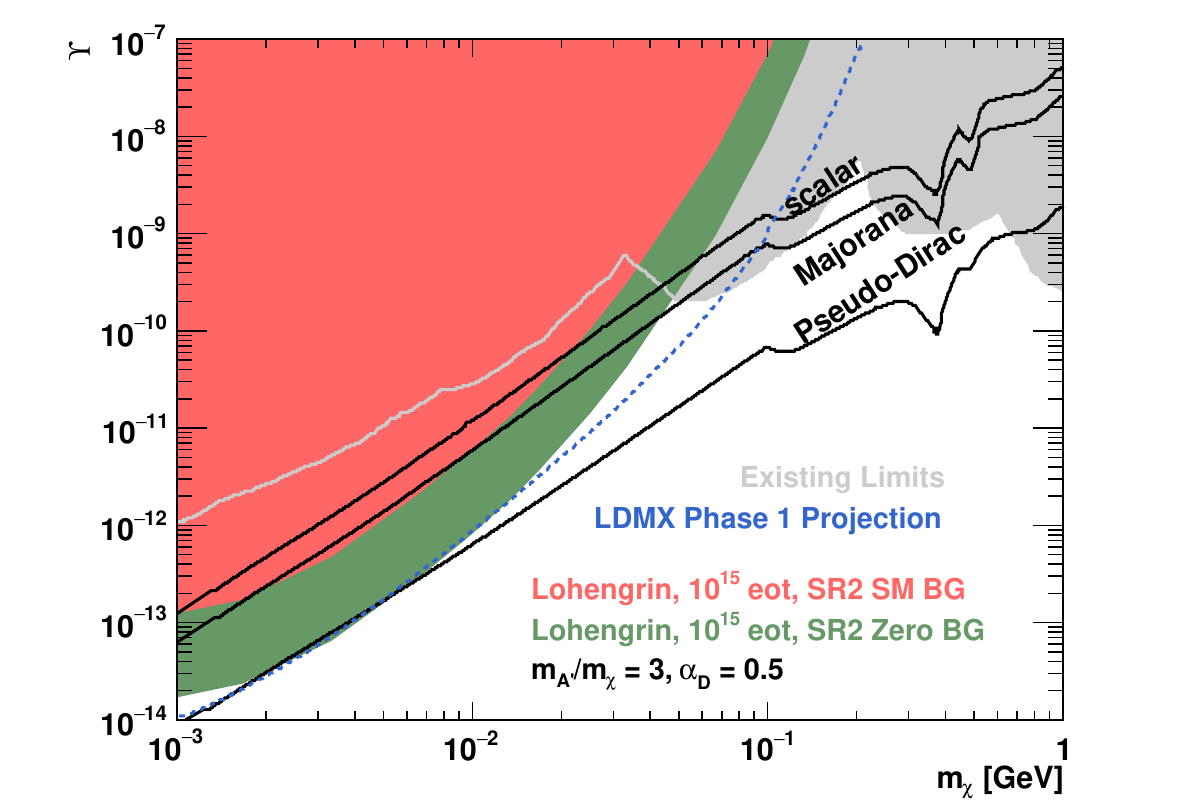}
\caption{Estimated sensitivity of the \lohengrin experiment for the extended ECAL coverage, an extended beam time and the more aggressive signal electron phase space. The red area shows the expected sensitivity including the estimated number of background events, and the green area indicates the statistical limit for the sensitivity assuming a background free search (both limits are calculated at the 90\% CL). The gray area shows existing limits from various experiments, in light blue the expected limit from Phase 1 of LDMX is shown.}
\label{fig:PhysicsReach:Sensitivity_SR2_LargerECAL}
\end{figure}

The expected sensitivity for the baseline scenario with the extended ECAL with the baseline signal region is shown in \cref{fig:PhysicsReach:Sensitivity_SR1_StandardECAL}. 
With 4$\times 10^{14}$ electrons on target, the \lohengrin experiment will be able to probe the dark sector parameter space just up to the expected properties for scalar dark matter, for $\SI{2}{\mega\electronvolt} \lesssim m_{\chi} \lesssim \SI{10}{\mega\electronvolt}$, roughly, assuming $m_{\ap}/m_{\chi} = 3$. Extending the ECAL coverage and choosing the more aggressive signal region has the potential to close the gap to Majorana dark matter for the same mass range, see Fig. \ref{fig:PhysicsReach:Sensitivity_SR2_LargerECAL}. If the expected backgrounds could be further reduced, e.g. by increasing the ECAL coverage, \lohengrin has the potential to close the gap all the way to Pseudo-Dirac dark matter, possibly with a slightly longer run time. For comparison the expected limit for Phase 1 of LDMX is shown as well. One major difference between \lohengrin and LDMX is the higher beam energy in LDMX. For the same number of electrons on target, this yields a substantially larger dark photon phase space that LDMX is sensitive to (see Fig. \ref{fig:PhysicsReach:Sensitivity_SR1_StandardECAL}). Assuming that the background suppression can be significantly improved compared to the baseline scenario that is assumed in this work, and that in addition the lower threshold for efficient electron reconstruction can be significantly lowered, the \lohengrin experiment could achieve a similar sensitivity as Phase 1 LDMX with a moderate increase in the recorded luminosity. We point out that expected sensitivity for the DarkSHINE experiment with $3\cdot10^{14}$ electrons on target is comparable to the LDMX Phase 1 sensitivity~\cite{chen2024darkshinebaselinedesignreport, Chen:2022liu}.

\begin{figure}[ht!]
\centering
\includegraphics[width=0.45\textwidth]{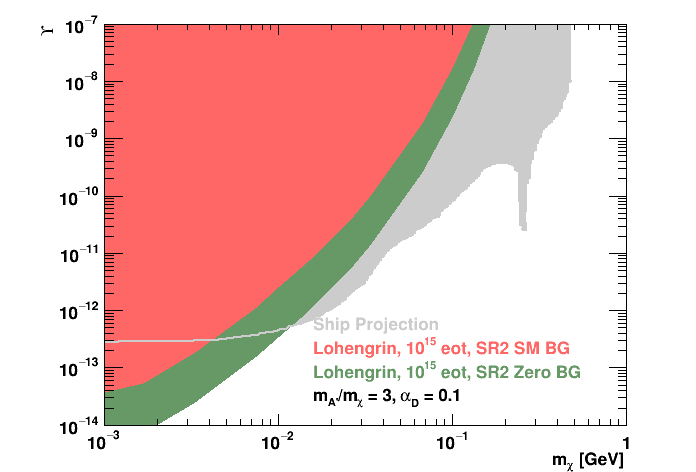}
\caption{Estimated sensitivity of the \lohengrin experiment for the extended ECAL coverage, an extended beam time and the more aggressive signal electron phase space. The red area shows the expected sensitivity including the estimated number of background events, and the green area indicates the statistical limit for the sensitivity assuming a background free search (both limits are calculated at the 90\% CL). The gray area shows the projected sensitivity of the SHiP experiment. In contrast to Figure \ref{fig:PhysicsReach:Sensitivity_SR1_StandardECAL} and Figure \ref{fig:PhysicsReach:Sensitivity_SR2_LargerECAL}, in this Figure a value of $\alpha_D = 0.1$ is assumed.}
\label{fig:PhysicsReach:Sensitivity_vs_Ship}
\end{figure}

While the relic targets move for varying $\alpha_D$, it is worth to note that the sensitivity of the Lohengrin experiment depends only on $\varepsilon$ and not on $\alpha_D$. The expected Lohengrin limits in $\upsilon$ hence scale with $\alpha_D$. This is an important difference to other planned experiments like the SHiP experiment. The SHiP Collaboration has published expected limits for $\alpha_D = 0.1$~\cite{Ahdida:2743145} - a comparison of the expected sensitivities for Lohengrin, using the extended ECAL coverage, the more aggresive signal region and an extended beam time, and SHiP are shown in Figure \ref{fig:PhysicsReach:Sensitivity_vs_Ship}. For dark photon masses below $~30$\,MeV the Lohengrin experiment might outperform the SHiP searches, provided that the background rejection can be improved compared to the assumptions that are made in this feasibility study.

%% file: roadmap.tex
A dark photon search experiment at the ELSA accelerator that has the potential to push the limits in the dark photon parameter space close to or beyond the relic targets is feasible. The realisation of such an experiment can proceed in parallel in several areas in multiple steps and phases:

First is the development of suitable front-end electronics for the tracking detector and the calorimeter. The analog signal processing in the readout ASICs must be made sufficiently fast in order to avoid pile-up induced deadtime and fake (veto) signals. Faster analog processing is possible, but care must be taken to maintain low noise levels and a reasonable resolution for all detectors.

In parallel to the development of suitable ASICs, a low rate pilot run can be performed using the current generation detectors. If the extraction rate is lowered by a few orders of magnitude, a tracking detector built from existing DMAPS and a prototype SiW calorimeter could be used to validate the modelling of crucial backgrounds. While events with SM bremsstrahlung photons that miss the ECAL seems to be well understood, events with electro-nuclear or photo-nuclear interactions in the target and/or the detectors are probably poorly modelled in \Geant, in particular for a large energy loss of the electron in the target. Low rate measurements can be used to solidify the confidence in the expectation values for these backgrounds, or help to improve the detector design to efficiently veto them, if required.

The \lohengrin experiment itself could be built in several phases. In a first phase, high rate tests without a hadron calorimeter could be done in order to verify the predictions for SM bremsstrahlung and establish the (non)relevance of VCS (virtual Compton scattering) as a possible background source. In a second phase the rate of events with neutral hadrons in the final state could be measured by adding a hadronic calorimeter to the setup.
For the first physics run with $4\cdot10^{14}$ electrons on target within a year, all detector components would be installed in their final destination.

%% file: conclusion.tex
We have presented a feasibility study for the \lohengrin experiment at the ELSA accelerator in Bonn, a high rate fixed target experiment that searches for new particles that feebly couple to the EM part of the SM.

The realisation of the proposed experiment, while challenging, is possible with some R\&D work required on the analog signal processing for the tracking detector and the electromagnetic calorimeter. Commissioning of the experiment would proceed in several phases, with each step subsequently consolidating the confidence in the physics potential of the experiment.

If successful, \lohengrin can close the gap between existing limits and the relic target for different dark matter models under the assumption of a dark photon that kinetically mixes with the SM photon (or hypercharge boson). Due to its sensitivity stemming from the \emph{disappearance} of momentum/energy, \lohengrin would be sensitive to extensions of the SM that go beyond the dark sector models that are used as a benchmark here.

\section*{Acknowledgements}
We are grateful for discussions with Pierre Thonet on matters regarding the magnet setup and Frank Frommberger on matters regarding the ELSA beam. We thank ELSA and TRA Matter of the University of Bonn for indispensable support. Funded by the Federal Ministry of Education and Research (BMBF) and the Ministry of Culture and Science of the State of North Rhine-Westphalia (MWK) as part of TRA Matter and the Excellence Strategy of the federal and state governments.